\documentclass[aps,11pt]{revtex4}
\usepackage{dcolumn}
\usepackage{graphicx}
\usepackage{amsmath}
\usepackage{amsfonts}
\usepackage{amssymb}
\usepackage{psfrag}
\usepackage{wrapfig}
\usepackage{subfigure}
\usepackage{makeidx}
\usepackage{bm}
\usepackage{epsf}
\usepackage{hyperref}
 \usepackage{float}
\usepackage{color}
\usepackage{cases}
\usepackage{booktabs} %绘制三线表
\usepackage{makecell} %表格内换行
\usepackage{multirow}
\usepackage{bm}%加粗 

\makeatletter

\newcommand{\Rmnum}[1]{\uppercase\expandafter{\romannumeral #1}}
\usepackage{amsmath} 
\begin{document} 

\title{Quasinormal modes of massless scalar and electromagnetic perturbations for Euler–Heisenberg black holes surrounded by perfect fluid dark matter}

\author{Chengfu Feng$^{1}$\footnote{fcfgyf@126.com}, Sheng-Yuan Li$^{1}$\footnote{shengyuanli77@outlook.com},
Xufen Zhang$^{1}$\footnote{xfzhangyzu@126.com},
Ming Zhang$^{2}$\footnote{mingzhang0807@126.com}, \\  
De-Cheng Zou$^{3}$\footnote{dczou@jxnu.edu.cn}
and Rui-Hong Yue$^{1}$\footnote{Corresponding author: rhyue@yzu.edu.cn}}

\address{
$^{1}$Center for Gravitation and Cosmology, College of Physical Science and Technology, Yangzhou University, Yangzhou 225009, China \\
$^{2}$Faculty of Science, Xihang University, Xi'an 710077 China \\
$^{3}$School of Physics, Jiangxi Normal University, Nanchang 330022, China}

\date{\today}

\begin{abstract}
\indent

We investigate the quasinormal modes of massless scalar and electromagnetic perturbations in charged Euler--Heisenberg black holes surrounded by perfect fluid dark matter. The quasinormal frequencies are calculated using the asymptotic iteration method and the sixth-order WKB approximation, and the relative deviation between the two methods is quantitatively analyzed to verify the reliability of results. The greybody factors for both perturbations are also evaluated within the sixth-order WKB framework. We systematically examine the effects of the black hole charge $Q$, nonlinear electrodynamic parameter $a$, dark matter parameter $\lambda$, and angular quantum number $l$ on the quasinormal frequencies and greybody factors. We find that these parameters significantly modify the structure of the effective potential barriers, and thus affect the oscillation frequencies, damping rates, and wave transmission and reflection properties of the perturbed fields.

\end{abstract}

%\keywords{}

\maketitle

\section{Introduction}
\label{intro}

The detection of gravitational waves from compact binary mergers by the LIGO and Virgo collaborations~\cite{Abbott:2016blz,Abbott:2017oio} has opened a new observational window into the strong-field regime of gravity. In particular, the ringdown phase of a perturbed black hole is governed by quasinormal modes (QNMs), which characterize the damped oscillations of spacetime under external perturbations~\cite{Kokkotas:1999bd,Berti:2009kk,Konoplya:2011qq}. Since the quasinormal spectrum depends on the parameters of the black hole and the underlying gravitational theory, QNMs provide powerful probes of black hole physics and possible deviations from general relativity~\cite{Cardoso:2019rvt,Berti:2018vdi}.

Meanwhile, dark matter remains one of the most important open problems in modern cosmology and astrophysics. Observations from galactic rotation curves~\cite{Rubin:1980zd}, galaxy cluster dynamics~\cite{Clowe:2006eq}, and cosmic microwave background measurements~\cite{Aghanim:2018eyx} provide compelling evidence for its existence. Among various phenomenological models, perfect fluid dark matter (PFDM) offers an effective framework for describing dark matter environments around black holes~\cite{Li:2013fka,Xu:2018, Konoplya:2025ect,Konoplya:2022hbl}. In this model, the surrounding dark matter distribution introduces logarithmic corrections to the spacetime geometry, thereby modifying particle motion, effective potentials, and wave propagation near black holes.

On the other hand, nonlinear electrodynamics plays an important role in strong electromagnetic field regimes. The Euler--Heisenberg (EH) effective action, derived from one-loop quantum electrodynamics~\cite{Euler:1935zz,Heisenberg:1936nmg}, describes vacuum polarization effects and naturally extends classical Maxwell theory. Black holes in nonlinear electrodynamics exhibit rich physical properties, including modified horizon structures, thermodynamics, optical behavior, and perturbation spectra~\cite{Kruglov:2017fut,Rodrigues:2023}. Recently, Euler--Heisenberg black holes surrounded by PFDM have attracted considerable attention~\cite{Ma:2024,Su:2024}. Such spacetimes incorporate both one-loop QED corrections characterized by the EH parameter $a$ and logarithmic dark matter corrections governed by the PFDM parameter $\lambda$, leading to rich phenomenological structures.

Recently, Belchior \textit{et al.}~\cite{Belchior:2026} investigated the greybody factors and absorption cross sections of spin-$0$, $1$, and $1/2$ fields propagating in Euler--Heisenberg black holes surrounded by PFDM. Using the Boonserm--Visser rigorous bound method, they showed that both the EH parameter and the PFDM parameter significantly modify the effective potential barrier and transmission probability. In addition, optical properties and shadow structures of black holes in dark matter environments have also been extensively studied in recent years~\cite{Akiyama:2019cqa,Akiyama:2022}.

Despite these developments, several important issues remain insufficiently explored. First, while greybody factors describe the scattering properties of perturbation fields, quasinormal modes characterize the intrinsic dynamical response of black holes under perturbations and dominate the ringdown stage of gravitational-wave signals. Second, most existing analyses in Euler--Heisenberg black hole spacetimes are restricted to the eikonal approximation or specific perturbation sectors, whereas systematic investigations of the complete quasinormal spectrum, including low multipoles and finite-$\ell$ effects, are still lacking. Third, asymptotically flat black holes remain of particular astrophysical relevance because realistic black holes such as M87$^*$~\cite{Akiyama:2019cqa} and Sagittarius A$^*$~\cite{Akiyama:2022} can be effectively described within asymptotically flat backgrounds.

Motivated by these considerations, in this work we systematically investigate the quasinormal modes and greybody factors of massless scalar and electromagnetic perturbations around asymptotically flat Euler--Heisenberg black holes surrounded by PFDM. The quasinormal frequencies are computed using the asymptotic iteration method (AIM)~\cite{Cho:2009cj} together with the sixth-order WKB approximation~\cite{Konoplya:2003ii} for comparison and consistency checks. We analyze the effects of the black hole charge $Q$, the Euler--Heisenberg parameter $a$, and the PFDM parameter $\lambda$ on the oscillation frequencies and damping rates. Our results reveal characteristic signatures of nonlinear electrodynamics and dark matter effects in black hole ringdown signals, providing potential theoretical predictions for future gravitational-wave observations.

This paper is organized as follows. In Sec.~II, we present the Euler--Heisenberg black hole spacetime in perfect fluid dark matter and derive the perturbation equations for scalar and electromagnetic fields, along with the corresponding effective potentials. In Sec.~III, we calculate the fundamental quasinormal frequencies via the asymptotic iteration method and sixth-order WKB approximation, and analyze their dependence on the charge $Q$, nonlinear parameter $a$, dark matter parameter $\lambda$, and angular momentum $l$. In Sec.~IV, we evaluate the greybody factors using the WKB method. Finally, we conclude and discuss the observational implications in Sec.~V. The numerical methods are detailed in the Appendices.

\section{Black Hole Spacetime and Perturbation Equations}

\subsection{ Background solution}

The spacetime geometry of an EH black hole surrounded by PFDM, originally constructed in Refs.~\cite{Ma:2024,Su:2024}, is described by the static and spherically symmetric line element
\begin{equation}
ds^2=-f(r)dt^2+\frac{1}{f(r)}dr^2+r^2\left(d\theta^2+\sin^2\theta\,d\phi^2\right),\label{metric}
\end{equation}
where the metric function \(f(r)\) takes the form~\cite{Ma:2024,Su:2024}
\begin{equation}
f(r)=1-\frac{2M}{r}+\frac{Q^2}{r^2}
-\frac{aQ^4}{20r^6}
+\frac{\lambda}{r}\ln\left|\frac{r}{\lambda}\right|.
\end{equation}

Here, \(M\) and \(Q\) denote the black hole mass and magnetic charge, respectively, while the parameter \(a\) characterizes the nonlinear corrections arising from the Euler--Heisenberg electrodynamics~\cite{Euler:1935zz,Heisenberg:1936nmg,Kruglov:2017fut}. The PFDM effect is encoded by the parameter \(\lambda\), which measures the intensity of the surrounding dark matter distribution~\cite{Li:2013fka,Xu:2018}.

Several well-known black hole solutions can be recovered in particular limits of the parameters:
\begin{itemize}
    \item For \(\lambda \to 0\), the solution reduces to the standard Euler--Heisenberg (EH) black hole spacetime~\cite{Ma:2024,Su:2024}.
    
    \item For \(Q \to 0\), the spacetime degenerates into a Schwarzschild black hole surrounded by PFDM~\cite{Li:2013,Xu:2018}.
    
    \item For both \(Q \to 0\) and \(\lambda \to 0\), the metric further reduces to the Schwarzschild black hole solution~\cite{Schwarzschild:1916}.
    
    \item In the combined limits \(a \to 0\) and \(\lambda \to 0\), the metric reduces to the Reissner--Nordstr\"om black hole solution~\cite{Reissner:1916,Nordstrom:1918}.

\end{itemize}

\subsection{ Scalar field perturbation}

Treating the scalar field as a test perturbation propagating in a fixed black hole background, its dynamics is governed by the massless Klein--Gordon equation
\begin{eqnarray}   
\Box\Phi
=\frac{1}{\sqrt{-g}}
\partial_{\mu}
\left(
\sqrt{-g}g^{\mu\nu}\partial_{\nu}\Phi
\right)
=0.
\label{eqscalar}
\end{eqnarray}  

By substituting the metric \eqref{metric} into Eq.~\eqref{eqscalar}, the scalar field can be decomposed through the method of separation of variables as~\cite{Regge:1957td}
\begin{eqnarray}  
\Phi(t,r,\theta,\varphi)
=
e^{-i\omega t}
\frac{\psi_s(r)}{r}
Y_{lm}(\theta,\varphi),
\end{eqnarray}  
where $Y_{lm}(\theta,\varphi)$ denotes the spherical harmonic function. After separating the angular variables, the radial part of the perturbation equation can be rewritten in the Schr\"odinger-like form
\begin{eqnarray} 
\frac{d^2 \psi_s(r_*)}{dr_*^{2}}
+
\left[
\omega^2-V_s(r)
\right]
\psi_s(r_*)
=0,
\label{E}
\end{eqnarray} 
where the tortoise coordinate $r_*$ is defined by
\begin{eqnarray} 
dr_*=\frac{dr}{f(r)}.
\label{rstar}
\end{eqnarray}

Here, $\psi_s(r)$ represents the radial wave function and $\omega$ corresponds to the quasinormal frequency. The effective potential $V_s(r)$ takes the form~\cite{Regge:1957td}
\begin{eqnarray}  
V_s(r)
=
\frac{
r f(r) f'(r)
+
f(r)l(l+1)
}{r^2},
\label{potVs} 
\end{eqnarray}
where $l$ is the multipole number of the quasinormal modes. It can be seen that the effective potential depends on both the black hole spacetime geometry and the multipole number $l$.

Fig.~1 displays the effective potential $V_{s}(r)$ for massless scalar perturbations around the Euler--Heisenberg black hole surrounded by perfect fluid dark matter with $M=1$. Panels (a) and (d) show that the potential barrier increases monotonically with the black hole charge $Q$, becoming higher and slightly narrower, with the PFDM background ($\lambda=0.1$) consistently producing a larger barrier than the vacuum case. Panels (b) and (e) indicate that the barrier height increases with the Euler--Heisenberg parameter $a$, although the effect remains relatively weak and the overall profile changes only slightly. Panels (c) and (f) demonstrate that increasing $l$ significantly raises and narrows the barrier, together with a small outward shift of the peak position, confirming the dominant role of the angular momentum term in the scalar effective potential. Panel (g) reveals a nonmonotonic dependence on $\lambda$: the barrier becomes higher and narrower at small and intermediate $\lambda$, while sufficiently large $\lambda$ leads to a lower and broader barrier. These results indicate that $l$, $Q$, and $\lambda$ primarily govern the scalar effective potential, whereas the nonlinear parameter $a$ introduces only mild corrections near the horizon.

\begin{figure}[htb]
\centering
\subfigure[$a=0.5$,$\lambda=0.1$ ,$l=0$]
{\label{fig11} %% label for first subfigure
\includegraphics[width=1.8in]{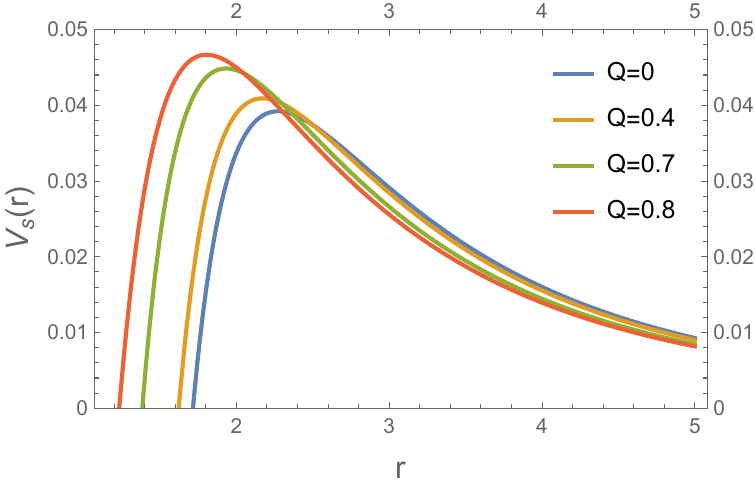}}
\subfigure[$Q=0.4$,$\lambda=0.1$ ,$l=0$]
{\label{fig13}%% label for first subfigure
\includegraphics[width=1.8in]{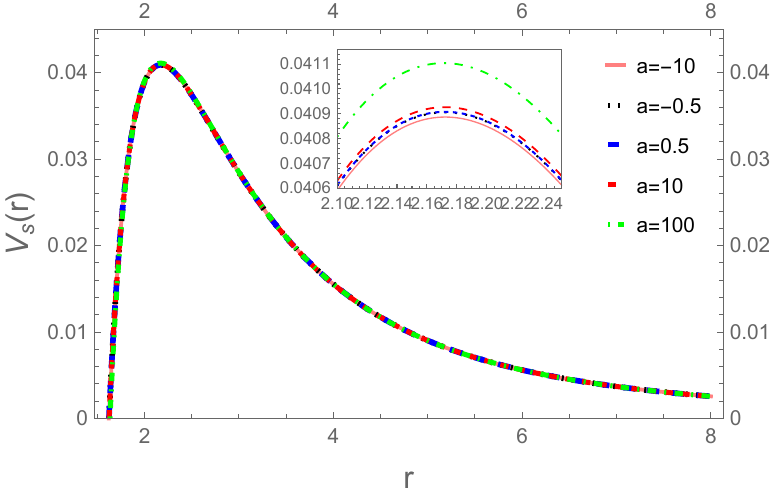}}
\subfigure[$a=0.5$,$\lambda=0.1$]
{\label{fig41} %% label for first subfigure
\includegraphics[width=1.8in]{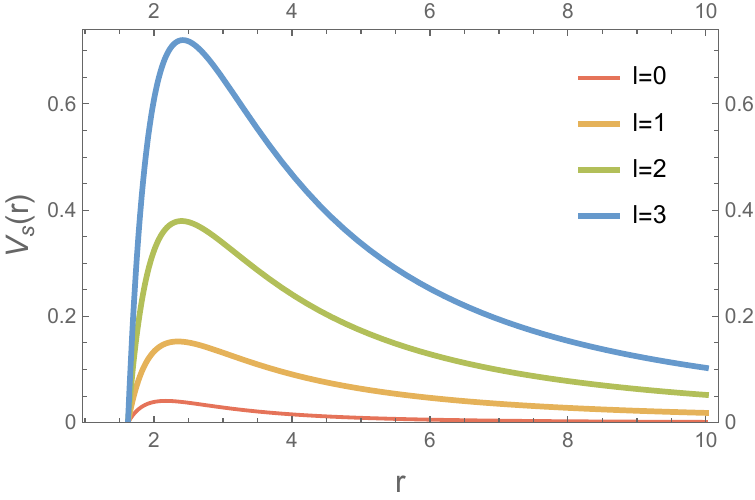}}
\subfigure[$a=0.5$,$\lambda=0$ ,$l=0$]
{\label{fig11} %% label for first subfigure
\includegraphics[width=1.8in]{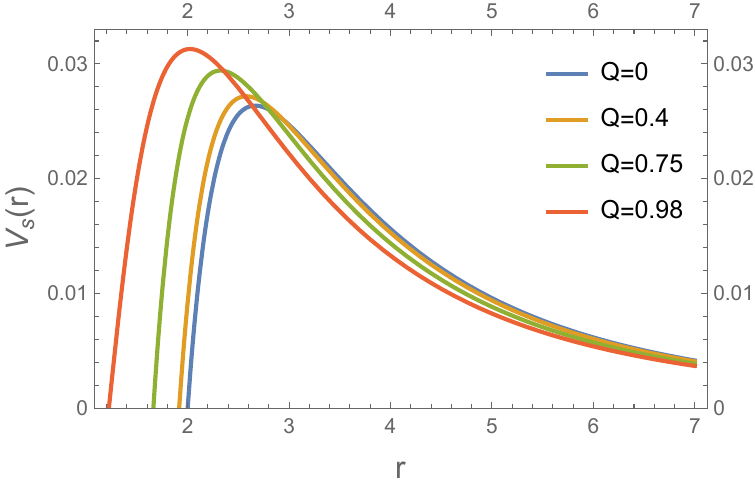}}
\subfigure[$Q=0.4$,$\lambda=0$ ,$l=0$]
{\label{fig13}%% label for first subfigure
\includegraphics[width=1.8in]{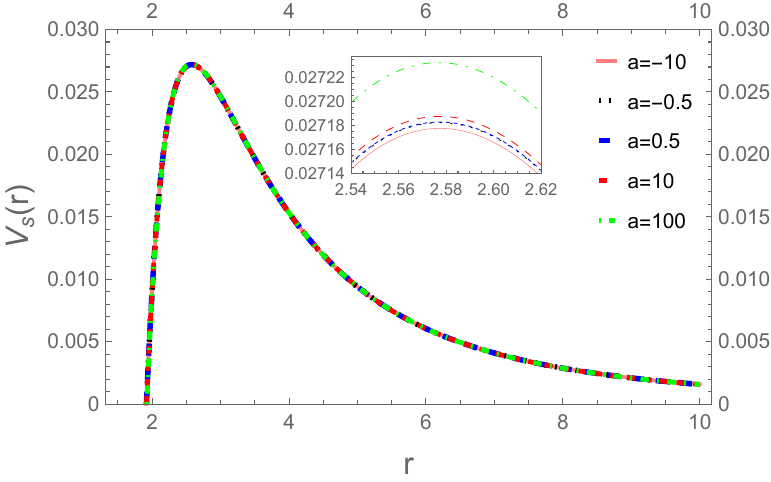}}
\subfigure[$a=0.5$,$\lambda=0$]
{\label{fig41} %% label for first subfigure
\includegraphics[width=1.8in]{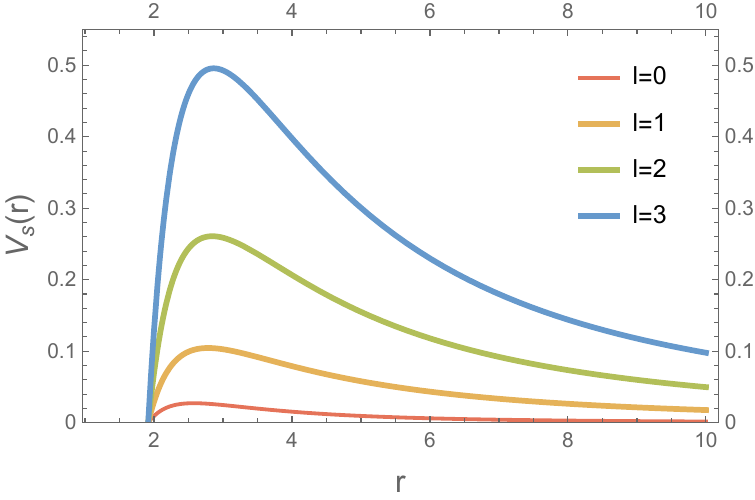}}
\subfigure[$a=0.5$,$Q=0.4$,$l=0$ ]
{\label{fig12} %% label for first subfigure
\includegraphics[width=1.8in]{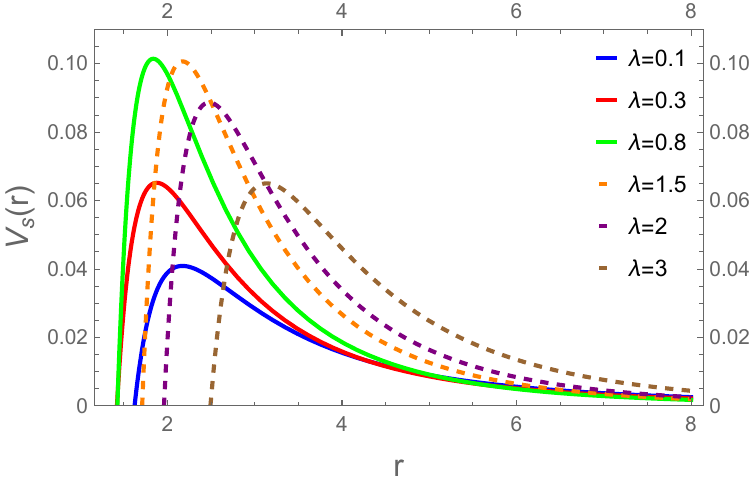}}
\caption{\textbf{The effective potential $V_s(r)$ for massless scalar field perturbation on black hole with $M=1$ }.}\label{fig1}
\end{figure}

\subsection{Electromagnetic field perturbation}

The dynamics of a massless electromagnetic field propagating in a curved spacetime and minimally coupled to the background geometry is governed by Maxwell’s equations
\begin{eqnarray}
\nabla_\mu F^{\mu\nu}
=
\frac{1}{\sqrt{-g}}
\partial_\mu
\left(
\sqrt{-g}\,F^{\mu\nu}
\right)
=0,
\label{eqmax}
\end{eqnarray}
where the electromagnetic field strength tensor is defined as
\begin{eqnarray}
F_{\mu\nu}
=
\partial_\mu A_\nu
-
\partial_\nu A_\mu,
\end{eqnarray}
with $A_\mu$ denoting the vector potential of the perturbed electromagnetic field. Following the standard decomposition~\cite{Regge:1957td,Vishveshwara:1970zz} in spherical symmetry, the vector potential can be expressed as
\begin{eqnarray}
A_\mu(t,r,\theta,\varphi)
=
\sum_{l,m}
e^{-i\omega t}
\begin{bmatrix}
0 \\
0 \\
h_0(r)\dfrac{1}{\sin\theta}
\dfrac{\partial Y_{l,m}}{\partial\varphi} \\
-h_0(r)\sin\theta
\dfrac{\partial Y_{l,m}}{\partial\theta}
\end{bmatrix}
+
\sum_{l,m}
e^{-i\omega t}
\begin{bmatrix}
h_1(r)Y_{l,m} \\
h_2(r)Y_{l,m} \\
h_3(r)\dfrac{\partial Y_{l,m}}{\partial\theta} \\
h_3(r)\dfrac{\partial Y_{l,m}}{\partial\varphi}
\end{bmatrix},
\label{eqA}
\end{eqnarray}
where $Y_{l,m}(\theta,\varphi)$ are the spherical harmonic functions, while $l$ and $m$ denote the angular and azimuthal quantum numbers, respectively. The first term in Eq.~\eqref{eqA} corresponds to the axial (odd-parity) sector with parity $(-1)^{l+1}$, whereas the second term represents the polar (even-parity) sector with parity $(-1)^l$.

Substituting the above decomposition into Maxwell’s equations \eqref{eqmax}, both the axial and polar electromagnetic perturbations can be reduced to a unified Schr\"odinger-like wave equation,
\begin{eqnarray}
\frac{d^2\Psi_e(r_*)}{dr_*^2}
+
\left[
\omega^2-V_e(r)
\right]
\Psi_e(r_*)
=0,
\label{e}
\end{eqnarray}
where the effective potential~\cite{Regge:1957td,Vishveshwara:1970zz} is given by
\begin{eqnarray}
V_e(r)
=
f(r)\frac{l(l+1)}{r^2}.
\label{potVe}
\end{eqnarray}

The master variable $\Psi_e(r)$ takes different forms for the odd- and even-parity sectors:
\begin{eqnarray}
&&\Psi_e(r)=h_0(r),
\qquad\qquad\qquad\textit{odd parity},
\\
&&\Psi_e(r)
=
-\frac{r^2}{l(l+1)}
\left(
i\omega h_2(r)
+
\frac{dh_1(r)}{dr}
\right),
\qquad
\textit{even parity}.
\end{eqnarray}

Fig.~2 displays the effective potential $V_{e}(r)$ for massless electromagnetic perturbations around the Euler--Heisenberg black hole surrounded by perfect fluid dark matter with $M=1$. Panels (a) and (d) show that the potential barrier increases with the black hole charge $Q$, with the enhancement becoming more significant near the extremal regime, similar to the scalar case. Panels (b) and (e) indicate that the barrier height decreases slightly with the Euler--Heisenberg parameter $a$, although the overall potential profile remains nearly unchanged, reflecting the weak influence of the nonlinear correction on electromagnetic perturbations. Panels (c) and (f) demonstrate that increasing $l$ significantly raises and narrows the barrier, with a smoother and more regular variation than in scalar perturbations. Panel (g) reveals a nonmonotonic dependence on $\lambda$: the barrier becomes higher and narrower at small $\lambda$, while sufficiently large $\lambda$ leads to a lower and broader barrier. These results confirm that $l$, $Q$, and $\lambda$ mainly govern the electromagnetic effective potential, whereas the nonlinear parameter $a$ introduces only mild near-horizon corrections.

\begin{figure}[H]
\centering
\subfigure[$a=0.5$,$\lambda=0.1$,$l=1$]
{\label{fig31} %% label for first subfigure
\includegraphics[width=1.8in]{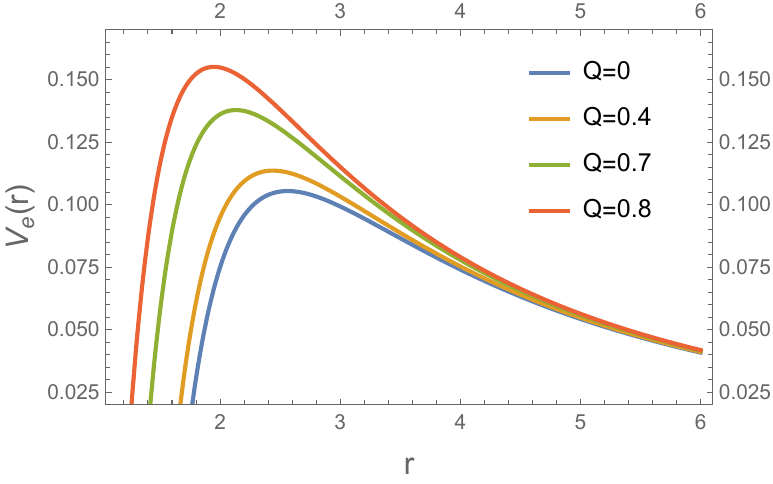}}
\subfigure[$Q=0.4$,$\lambda=0.1$,$l=1$]
{\label{fig33}%% label for first subfigure
\includegraphics[width=1.8in]{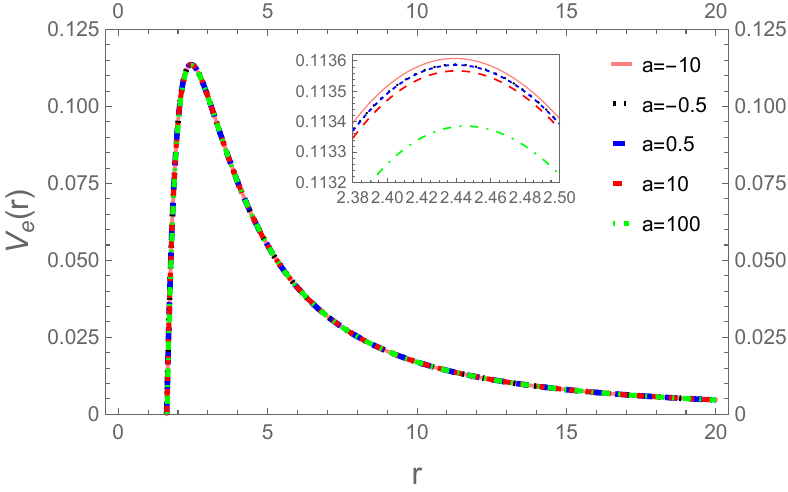}}
\subfigure[$a=0.5$,$\lambda=0.1$]
{\label{fig41} %% label for first subfigure
\includegraphics[width=1.8in]{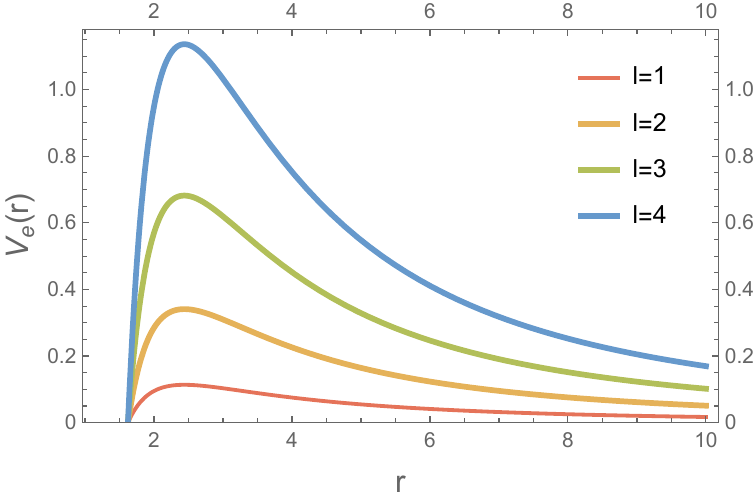}}
\subfigure[$a=0.5$,$\lambda=0$,$l=1$]
{\label{fig31} %% label for first subfigure
\includegraphics[width=1.8in]{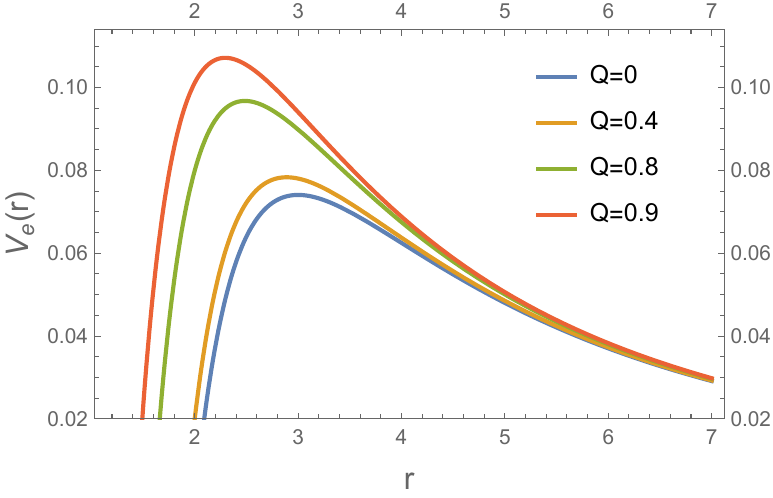}}
\subfigure[$Q=0.4$,$\lambda=0$,$l=1$]
{\label{fig33}%% label for first subfigure
\includegraphics[width=1.8in]{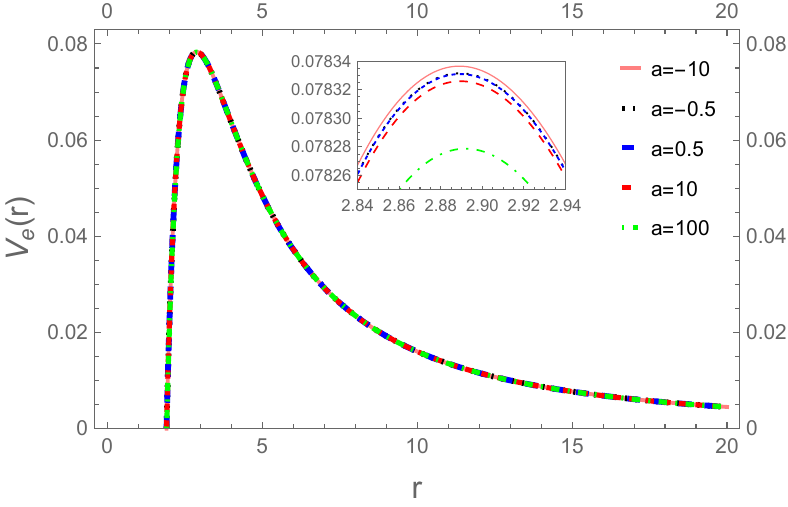}}
\subfigure[$a=0.5$,$\lambda=0$]
{\label{fig41} %% label for first subfigure
\includegraphics[width=1.8in]{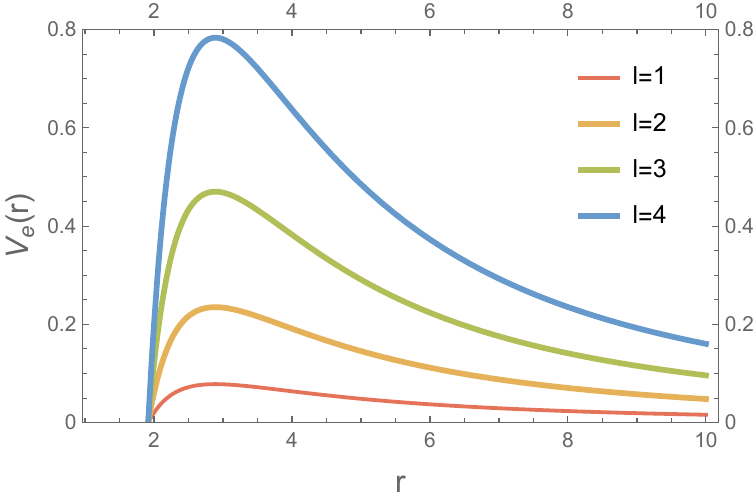}}
\subfigure[$a=0.5$,$Q=0.4$,$l=1$]
{\label{fig32} %% label for first subfigure
\includegraphics[width=1.8in]{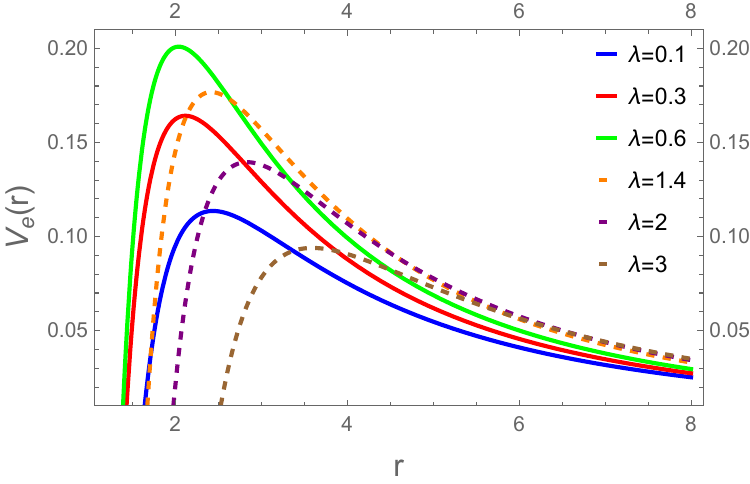}}
\caption{\textbf{The effective potential $V_e(r)$ for massless electromagnetic field on black hole with $M=1$ }.}\label{fig3}
\end{figure}

\section{Quasinormal Mode frequencies}
\label{sec3}

With the effective potentials constructed for scalar and electromagnetic perturbations in hand, we proceed to compute the quasinormal frequencies (QNFs) of magnetically charged Euler–Heisenberg black holes surrounded by perfect fluid dark matter. The quasinormal frequencies $\omega$ correspond to characteristic complex oscillation frequencies of the perturbed spacetime, subject to physically motivated boundary conditions~\cite{Konoplya:2011qq,Berti:2009kk} imposed on the radial wave function:
\begin{align}
\psi(r_{*}) &\sim e^{-i \omega r_{*}}, \quad r_{*} \to -\infty,\\
\psi(r_{*}) &\sim e^{i \omega r_{*}}, \quad r_{*} \to +\infty,
\end{align}
which demand purely ingoing wave solutions at the event horizon and purely outgoing wave solutions at spatial infinity.

Since the perturbation equations cannot be solved analytically in closed form, we employ two independent, high-precision numerical approaches to determine the fundamental quasinormal frequencies ($n=0$), which dominate the black hole ringdown signal in gravitational-wave observations. Specifically, we use the asymptotic iteration method (AIM)~\cite{Cho:2009cj} and the sixth-order Wentzel–Kramers–Brillouin (WKB) approximation~\cite{Konoplya:2003ii}. Detailed implementations of both methods are summarized in Appendices A and B for completeness.

To quantify the consistency between the two sets of results, we define the relative percentage deviation
\begin{align}
\Delta_{AW}=\frac{|\omega_{\text{AIM}}-\omega_{\text{WKB}}|}{|\omega_{\text{WKB}}|}\times 100\%,
\end{align}
which serves as a reliability check for our computed QNFs.

Our numerical results for the fundamental modes are systematically presented in Tables I and II, while the parametric dependences on the black hole charge $Q$, nonlinear parameter $a$, dark matter parameter $\lambda$, and angular momentum $l$ are illustrated in Figs.~3–14. Throughout the analysis, we concentrate on the fundamental mode $n=0$, since higher overtones ($n\geq1$) decay much faster and are generally not observable in current and upcoming gravitational-wave detectors.

\begin{table*}[h!]
\caption{The fundamental QNFs $(n=0)$ for scalar perturbation with $M=1$.}
\label{Table1}
\resizebox{\linewidth}{!}{
\begin{tabular}{|c|c|c|c|c|c|c|c|c|}
\hline
$a$ & $l$ & $Q$ 
& \multicolumn{3}{c|}{$\lambda=0.1$} 
& \multicolumn{3}{c|}{$\lambda=0$} \\
\cline{4-9}
& & & $\omega_{\text{AIM}}$ & $\omega_{\text{WKB}}$ & $\Delta_{\text{AW}}$ 
& $\omega_{\text{AIM}}$ & $\omega_{\text{WKB}}$ & $\Delta_{\text{AW}}$ \\
\hline

\multirow{12}{*}{$0.5$}
&\multirow{7}{*}{0}
&0.0&$0.134883626-0.126826778i$&$0.132684-0.124198i$&$1.886\%$&$0.112820904-0.102208173i$&$0.110493-0.100793i$&$1.82167\%$\\
&&0.1&$0.135209625-0.126912332i$&$0.13306-0.124238i$&$1.88472\%$&$0.113026236-0.102253336i$&$0.110631-0.100895i$&$1.839\%$\\
&&0.2&$0.136206308-0.127165291i$&$0.133893-0.124644i$&$1.87031\%$&$0.113651158-0.102386650i$&$0.11127-0.101001i$&$1.83308\%$\\
&&0.3&$0.137933044-0.127572332i$&$0.135531-0.125126i$&$1.85903\%$&$0.114723623-0.102600720i$&$0.112272-0.101252i$&$1.85084\%$\\
\cline{2-9}
&\multirow{7}{*}{1}
&0.0&$0.349762033-0.120008018i$&$0.349714-0.120175i$&$0.0471\%$&$0.292936749-0.097648398i$&$0.29291-0.0977616i$&$0.0377039\%$\\
&&0.1&$0.350542936-0.120100118i$&$0.350499-0.120268i$&$0.0469925\%$&$0.293435533-0.097703087i$&$0.293407-0.0978114i$&$0.0362537\%$\\
&&0.2&$0.352941458-0.120388924i$&$0.352898-0.120545i$&$0.0435261\%$&$0.294944839-0.097848831i$&$0.294919-0.0979589i$&$0.036349\%$\\
&&0.3&$0.357090901-0.120849311i$&$0.357056-0.120996i$&$0.0401007\%$&$0.297539532-0.098093302i$&$0.297516-0.0981976i$&$0.0341002\%$\\
\cline{2-9}
&\multirow{7}{*}{2}
&0.0&$0.577130589-0.118814142i$&$0.577127-0.118825i$&$0.00199031\%$&$0.483643303-0.096759690i$&$0.483642-0.0967661i$&$0.00132928\%$\\
&&0.1&$0.578411754-0.118911749i$&$0.578409-0.118923i$&$0.00198025\%$&$0.484456887-0.096811923i$&$0.484455-0.0968185i$&$0.00139392\%$\\
&&0.2&$0.582328875-0.119202703i$&$0.582325-0.119213i$&$0.00186565\%$&$0.486930976-0.096966645i$&$0.486929-0.0969739i$&$0.00149251\%$\\
&&0.3&$0.589114577-0.119676234i$&$0.589112-0.119687i$&$0.00180418\%$&$0.491181272-0.097220120i$&$0.491179-0.0972261i$&$0.00129818\%$\\
\hline

\multirow{12}{*}{$-0.5$}
&\multirow{7}{*}{0}
&0.0&$0.134883626-0.126826778i$&$0.132684-0.124198i$&$1.886\%$&$0.112820904-0.102208173i$&$0.110493-0.100793i$&$1.82167\%$\\
&&0.1&$0.135209638-0.126912307i$&$0.13298-0.124313i$&$1.88123\%$&$0.113026240-0.102253328i$&$0.110651-0.100877i$&$1.83356\%$\\
&&0.2&$0.136206523-0.127164867i$&$0.133897-0.124644i$&$1.86895\%$&$0.113651223-0.102386514i$&$0.111258-0.101013i$&$1.83612\%$\\
&&0.3&$0.137934210-0.127569891i$&$0.135439-0.125231i$&$1.85403\%$&$0.114723966-0.102599957i$&$0.112296-0.101237i$&$1.84171\%$\\
\cline{2-9}
&\multirow{7}{*}{1}
&0.0&$0.349762033-0.120008018i$&$0.349714-0.120175i$&$0.0471\%$&$0.292936749-0.097648398i$&$0.29291-0.0977616i$&$0.0377039\%$\\
&&0.1&$0.350543990-0.120102304i$&$0.350499-0.120268i$&$0.0464806\%$&$0.293434397-0.097698491i$&$0.293407-0.0978114i$&$0.0375942\%$\\
&&0.2&$0.352939388-0.120386933i$&$0.352899-0.120545i$&$0.0437566\%$&$0.294945412-0.097852198i$&$0.294919-0.0979587i$&$0.0352797\%$\\
&&0.3&$0.357090044-0.120846111i$&$0.357057-0.120994i$&$0.0401772\%$&$0.297539522-0.098095061i$&$0.297517-0.0981968i$&$0.0332767\%$\\
\cline{2-9}
&\multirow{7}{*}{2}
&0.0&$0.577130589-0.118814142i$&$0.577127-0.118825i$&$0.00199031\%$&$0.483643303-0.096759690i$&$0.483642-0.0967661i$&$0.00132928\%$\\
&&0.1&$0.578412103-0.118911979i$&$0.578409-0.118923i$&$0.00195617\%$&$0.484456220-0.096811781i$&$0.484455-0.0968185i$&$0.00138579\%$\\
&&0.2&$0.582328603-0.119201934i$&$0.582326-0.119213i$&$0.00191009\%$&$0.486931684-0.096967346i$&$0.486930-0.0969738i$&$0.00136761\%$\\
&&0.3&$0.589115528-0.119673888i$&$0.589113-0.119685i$&$0.00185381\%$&$0.491181408-0.097218829i$&$0.491179-0.0972255i$&$0.00140159\%$\\
\hline
\end{tabular}
}
\end{table*}

\begin{table*}[h!]
\caption{The fundamental QNFs $(n=0)$ for electromagnetic perturbation with $M=1$.}
\label{Table1}
\resizebox{\linewidth}{!}{
\begin{tabular}{|c|c|c|c|c|c|c|c|c|}
\hline
$a$ & $l$ & $Q$ 
& \multicolumn{3}{c|}{$\lambda=0.1$} 
& \multicolumn{3}{c|}{$\lambda=0$} \\
\cline{4-9}
& & & $\omega_{\text{AIM}}$ & $\omega_{\text{WKB}}$ & $\Delta_{\text{AW}}$ 
& $\omega_{\text{AIM}}$ & $\omega_{\text{WKB}}$ & $\Delta_{\text{AW}}$ \\
\hline

\multirow{12}{*}{$0.5$}
&\multirow{7}{*}{1}
&0.0&$0.29439285 - 0.11316717i$&$0.294266 - 0.113376 i$&$0.0776656\%$&$0.24828784 - 0.09248966i$&$0.248191 - 0.0926371 i$&$0.06652\%$\\
&&0.1&$0.29513482- 0.11327844i$&$0.295007 - 0.113491  i$&$0.0786329\%$&$0.24875470 - 0.09255345i$&$0.248666 - 0.0927002   i$&$0.0646035\%$\\
&&0.2&$0.29740663- 0.11361424i$&$0.297277 - 0.113835 i$&$0.0803215\%$&$0.25021011 - 0.09273381i$&$0.250111 - 0.092889 i$&$0.0689933\%$\\
&&0.3&$0.30135147 - 0.11416787i$&$0.301223 - 0.114402 i$&$0.0830399\%$&$0.25269744 - 0.09303902i$&$0.252599 - 0.093199 i$&$0.0697181\%$\\
\cline{2-9}
&\multirow{7}{*}{2}
&0.0&$0.54484126 - 0.11649221i$&$0.544838 - 0.116501  i$&$0.00173403\%$&$0.45759495 - 0.09500484i$&$0.457593 - 0.0950111 i$&$0.00138925\%$\\
&&0.1&$0.54609611 - 0.11659376i$&$0.546094 - 0.116604  i$&$0.00189084\%$&$0.45839448- 0.09506061i$&$0.458393 - 0.0950668 i$&$0.00139081\%$\\
&&0.2&$0.54993578 - 0.11690003i$&$0.549933 - 0.11691 i$&$0.00180058\%$&$0.46082780 - 0.09522534i$&$0.460826 - 0.0952323 i$&$0.00151945\%$\\
&&0.3&$0.55659497 - 0.11740034i$&$0.556592 - 0.117411  i$&$0.00188504\%$&$0.46500970 - 0.09549510i$&$0.465009 - 0.0955021  i$&$0.00148841\%$\\
\cline{2-9}
&\multirow{7}{*}{3}
&0.0&$0.78289802- 0.11729825i$&$0.782897 - 0.1173 i$&$0.000173145\%$&$0.65689875 - 0.09561645i$&$0.656898 - 0.0956171  i$&$0.000108951\%$\\
&&0.1&$0.78466270 - 0.11739866i$&$0.784663 - 0.1174 i$&$0.000179414\%$&$0.65802125 - 0.09567019i$&$0.658021 - 0.0956715 i$&$0.000203299\%$\\
&&0.2&$0.79006023- 0.11769832i$&$0.79006 - 0.1177  i$&$0.000160961\%$&$0.66143963- 0.09583222i$&$0.661439 - 0.0958331 i$&$0.000177894\%$\\
&&0.3&$0.79941657 - 0.11818833i$&$0.799417 - 0.11819 i$&$0.000157755\%$&$0.66731036- 0.09609459i$&$0.66731 - 0.0960961 i$&$0.000230448\%$\\
\hline

\multirow{21}{*}{$-0.5$}
&\multirow{7}{*}{1}
&0.0&$0.29439285 - 0.11316717i$&$0.294266 - 0.113376 i$&$0.0776656\%$&$0.24828784 - 0.09248966i$&$0.248191 - 0.0926371  i$&$0.06652\%$\\
&&0.1&$0.29513429 - 0.11327869i$&$0.295007 - 0.113491   i$&$0.0784021\%$&$0.24876447 - 0.09254923i$&$0.248666 - 0.0927003 i$&$0.0680151\%$\\
&&0.2&$0.29740642- 0.11361308i$&$0.297278 - 0.113834 i$&$0.0804621\%$&$0.25020484 - 0.09273661i$&$0.250111 - 0.0928888 i$&$0.066916\%$\\
&&0.3&$0.30135473 - 0.11416639i$&$0.301226 - 0.114399 i$&$0.082579\%$&$0.25269255- 0.09303844i$&$0.252601 - 0.0931979 i$&$0.0683725\%$\\
\cline{2-9}
&\multirow{7}{*}{2}
&0.0&$0.54484126 - 0.11649221i$&$0.544838 - 0.116501  i$&$0.00173403\%$&$0.45759495 - 0.09500484i$&$0.457593 - 0.0950111  i$&$0.00138925\%$\\
&&0.1&$0.54609676 - 0.11659470i$&$0.546094 - 0.116604 i$&$0.00175835\%$&$0.45839371 - 0.09506030i$&$0.458393 - 0.0950668 i$&$0.00141575\%$\\
&&0.2&$0.54993591 - 0.11689937i$&$0.549933 - 0.116909 i$&$0.00184227\%$&$0.46082878 - 0.09522548i$&$0.460826 - 0.0952322 i$&$0.00151643\%$\\
&&0.3&$0.55659606 - 0.11739805i$&$0.556594 - 0.117409 i$&$0.00189805\%$&$0.46501125- 0.09549540i$&$0.465009 - 0.0955014 i$&$0.00133511\%$\\
\cline{2-9}
&\multirow{7}{*}{3}
&0.0&$0.78289802 - 0.11729825i$&$0.782897 - 0.1173  i$&$0.000173145\%$&$0.65689875- 0.09561645i$&$0.656898 - 0.0956171 i$&$0.000108951\%$\\
&&0.1&$0.78466307- 0.11739848i$&$0.784663 - 0.1174 i$&$0.000208936\%$&$0.65802150- 0.09567039i$&$0.658021 - 0.0956715  i$&$0.000185181\%$\\
&&0.2&$0.79006028- 0.11769779i$&$0.79006 - 0.117699 i$&$0.000182509\%$&$0.66143867- 0.09583175i$&$0.661439 - 0.095833 i$&$0.00019171\%$\\
&&0.3&$0.79941958 - 0.11818638i$&$0.799419 - 0.118188 i$&$0.000196239\%$&$0.66731087 - 0.09609473i$&$0.667311 - 0.0960955 i$&$0.000117978\%$\\
\hline
\end{tabular}
}
\end{table*}

\subsection{\textbf{\texorpdfstring{Dependence on the angular quantum number $l$}{l--dependence}}}
Fig.3presents the fundamental quasinormal frequencies ($n=0$) for massless scalar and electromagnetic perturbations around the Euler--Heisenberg black hole surrounded by perfect fluid dark matter with $M=1$ and $Q=0.4$. The vertical axis represents the real part $\omega_R$ corresponding to the oscillation frequency, while the horizontal axis shows the imaginary part $\omega_I$ characterizing the damping rate. Arrows denote the direction of increasing angular quantum number $l$. We display four typical cases with fixed nonlinear parameters: (a) $a=0.5,\lambda=0.1$, (b) $a=-0.5,\lambda=0.1$, (c) $a=0.5,\lambda=0$, and (d) $a=-0.5,\lambda=0$.

The physical mechanism of the $l$-dependence can be rigorously traced back to the structure of the effective potential barriers presented in Sec. II. As illustrated in Fig.~1 for the scalar potential $V_s(r)$ and Fig.~2 for the electromagnetic potential $V_e(r)$, increasing $l$ universally leads to a higher and narrower potential barrier for both types of perturbations. Since the oscillation frequency $\omega_R$ is mainly governed by the height of the potential barrier, $\omega_R$ increases monotonically with $l$ for both scalar and electromagnetic perturbations, which reflects stronger confinement of the field fluctuations.

For the damping behavior described by $\omega_I$, scalar and electromagnetic perturbations exhibit opposite trends due to their distinct potential constructions. For scalar perturbations, $|\omega_I|$ decreases with increasing $l$, indicating that the narrower potential barrier weakens the decay of perturbations. For electromagnetic perturbations, $|\omega_I|$ increases with increasing $l$, meaning that the narrower potential barrier enhances the decay of perturbations. This distinction originates from the different dependences of the near-horizon curvature and barrier steepness on $l$ between $V_s(r)$ and $V_e(r)$.

Compared with the vacuum case $\lambda=0$, the presence of perfect fluid dark matter ($\lambda=0.1$) renders the effective potential barrier higher and narrower, which simultaneously enhances both $\omega_R$ and $|\omega_I|$. Moreover, the influence of dark matter becomes more significant at larger $l$, indicating that a higher angular momentum amplifies the effect of dark matter on perturbation dynamics.

Throughout this subsection, the nonlinear Euler--Heisenberg parameter is fixed at $a=0.5$ and $a=-0.5$, representing weak nonlinear corrections from quantum electrodynamics. This choice of $a$ only introduces mild modifications near the event horizon and barely affects the global profile of $V_s(r)$ and $V_e(r)$, thus yielding negligible effects on the $l$-dependent behavior of quasinormal frequencies. This confirms that the angular quantum number $l$ dominates the potential barrier structure and further determines the key characteristics of quasinormal modes.

\begin{figure}[H]
\centering
\subfigure[$a=0.5,\lambda=0.1$]
{%% label for first subfigure
\includegraphics[width=2.5in]{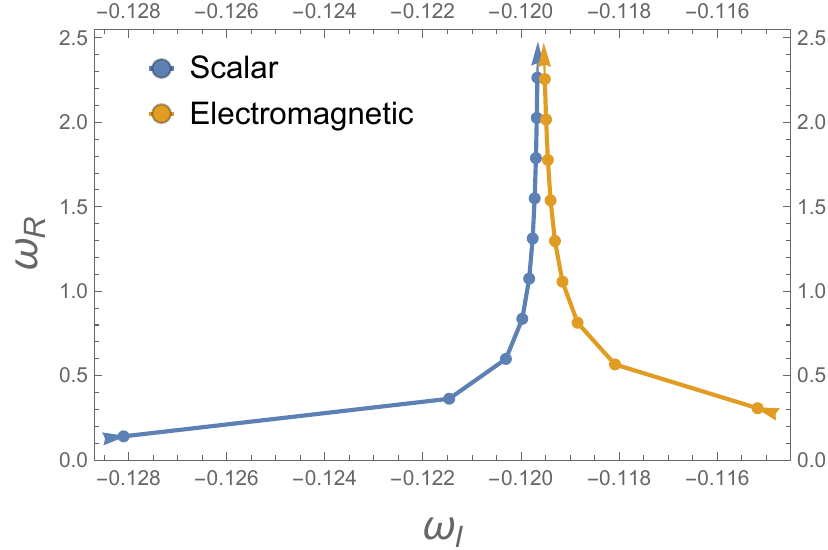}}
\subfigure[$a=-0.5,\lambda=0.1$]
{ %% label for first subfigure
\includegraphics[width=2.5in]{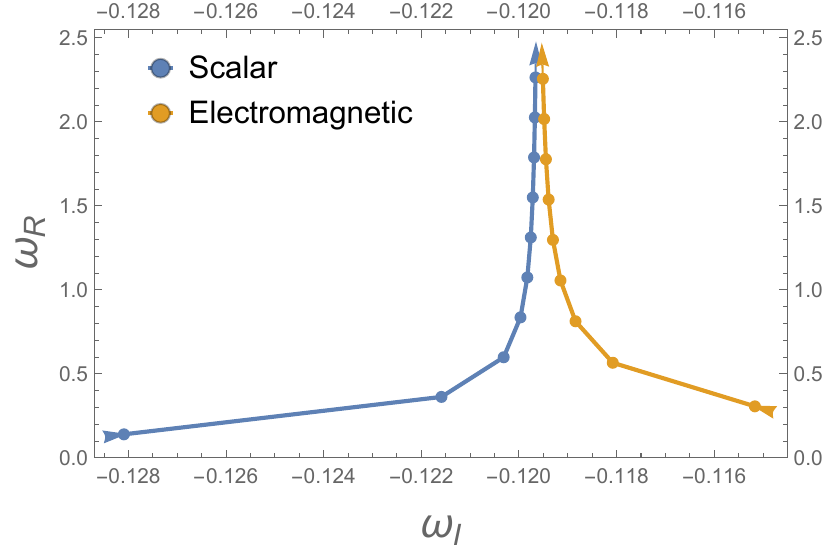}}
\subfigure[$a=0.5,\lambda=0$]
{%% label for first subfigure
\includegraphics[width=2.5in]{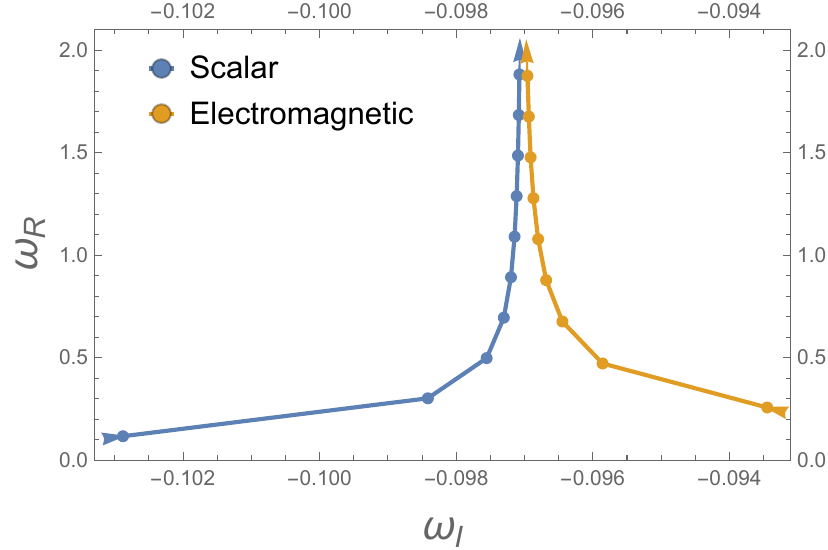}}
\subfigure[$a=-0.5,\lambda=0$]
{ %% label for first subfigure
\includegraphics[width=2.5in]{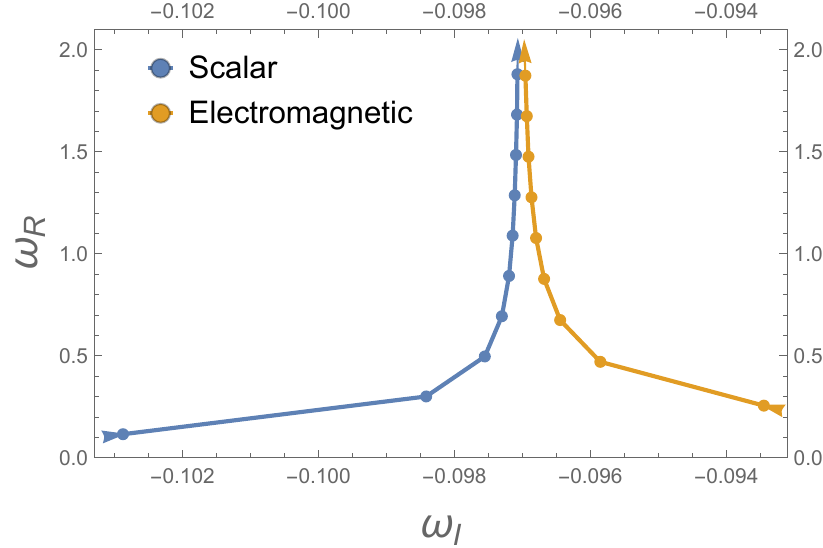}}
\caption{ Variation of scalar and electromagnetic fundamental QNFs  ($n=0$) with respect to different $l$ for $M=1$ and $Q=0.4$.The arrow indicates the increasing direction of the angular momentum parameter \(l\). Panels (a), (b), (c), and (d) show the scalar field modes with \(l=0, 1, 2, 3, 4, 5, 6, 7, 8\) and the electromagnetic field modes with \(l=1, 2, 3, 4, 5, 6, 7, 8\), respectively.}\label{figl1}
\end{figure}

\subsection{\textbf{\texorpdfstring{Dependence on the dark matter parameter $\lambda$}{λ--dependence}}}

In the background of Euler--Heisenberg black holes surrounded by perfect fluid dark matter with unit mass $M=1$, the fundamental quasinormal frequencies ($n=0$) of massless scalar and electromagnetic perturbations exhibit significant nonmonotonic behavior with the dark matter parameter $\lambda$, as illustrated in Figs.~4--7, where the arrows denote the direction of increasing $\lambda$, and the black and purple dots mark the maxima of $\omega_R$ and minima of $\omega_I$, respectively. For scalar perturbations, $\omega_R$ first increases and then decreases with $\lambda$, while $\omega_I$ first decreases and then increases, and the critical $\lambda$ corresponding to the extrema gradually decreases as the angular quantum number $l$ grows; for electromagnetic perturbations, similar nonmonotonic behavior appears, but the critical $\lambda$ for extrema increases with $l$, which arises from the extra derivative term in the scalar effective potential making scalar perturbations more sensitive to the near-horizon geometry modified by PFDM. The increase of black hole charge $Q$ shifts the critical $\lambda$ to smaller values for both types of perturbations, advancing the onset of the nonmonotonic behavior induced by $\lambda$. Meanwhile, there is a strong coupling between the nonlinear parameter $a$ and $\lambda$: increasing $a$ moves the critical $\lambda$ for the maximum of $\omega_R$ to larger values and the critical $\lambda$ for the minimum of $\omega_I$ to smaller values; even in the weak dark matter regime with small $\lambda$, the evolution curve of $\lambda$ is still significantly shifted and modulated by large $a$ (strong nonlinearity), with obvious changes in the extremum positions and variation ranges, and only in the strong dark matter region with sufficiently large $\lambda$ does the influence of $a$ gradually weaken and become dominated by the PFDM effect. Overall, $\lambda$ acts as the core parameter driving the nonmonotonic characteristics of the quasinormal mode spectrum, and its coupling with $Q$ and $a$ jointly shapes the complex oscillation and damping behaviors of the perturbation fields.

\begin{figure}[H]
\centering
\subfigure[scalar,$a=0.5$]
{\label{figeps11} %% label for first subfigure
\includegraphics[width=1.53in]{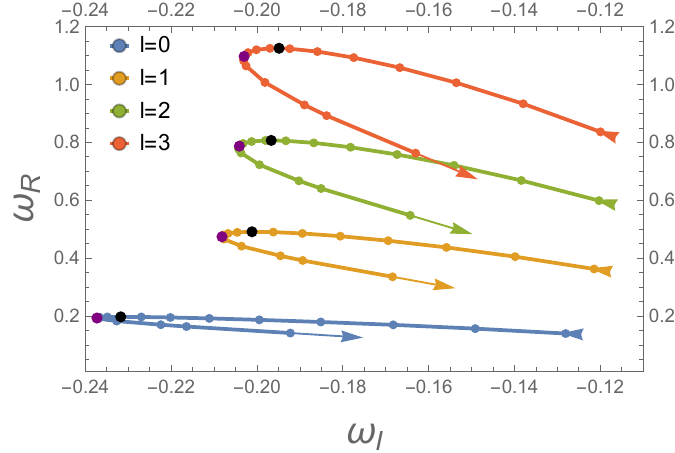}}
\subfigure[scalar,$a=-0.5$]
{\label{figeps12} %% label for first subfigure
\includegraphics[width=1.53in]{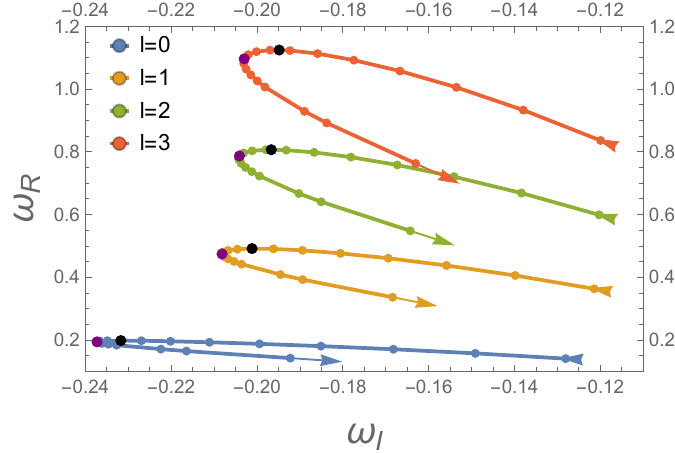}}
\subfigure[electromagnetic,$a=0.5$]
{\label{figeps11} %% label for first subfigure
\includegraphics[width=1.53in]{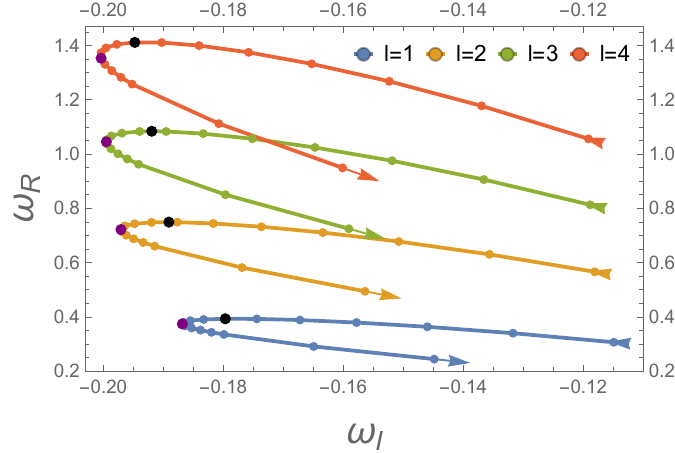}}
\subfigure[electromagnetic,$a=-0.5$]
{\label{figeps12} %% label for first subfigure
\includegraphics[width=1.55in]{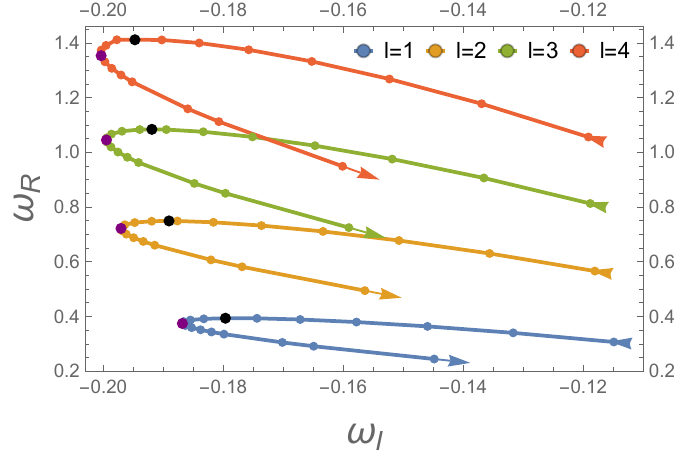}}
\caption{Variation of scalar and electromagnetic fundamental QNFs ($n=0$) with respect to different $\lambda$ for $M=1$ and $Q=0.4$. The arrows in panels (a)--(d) show the increasing direction of $\lambda$ with $\lambda\in[0.1,3]$. Black dots denote the maxima of $\omega_R$ and purple dots denote the minima of $\omega_I$ versus $\lambda$. For scalar QNFs with $l=0,1,2,3$, the black dots are near $\lambda=0.9,0.8,0.77,0.75$ and purple dots near $\lambda=1.22,1.19,1.17,1.15$; for electromagnetic QNFs with $l=1,2,3,4$, the black dots are near $\lambda=0.7,0.73,0.75,0.8$ and purple dots near $\lambda=1.1,1.13,1.15,1.2$.
}\label{figeps1}
\end{figure}

\begin{figure}[H]
\centering
\subfigure[$a=0.5$]
{\label{figeps11} %% label for first subfigure
\includegraphics[width=1.7in]{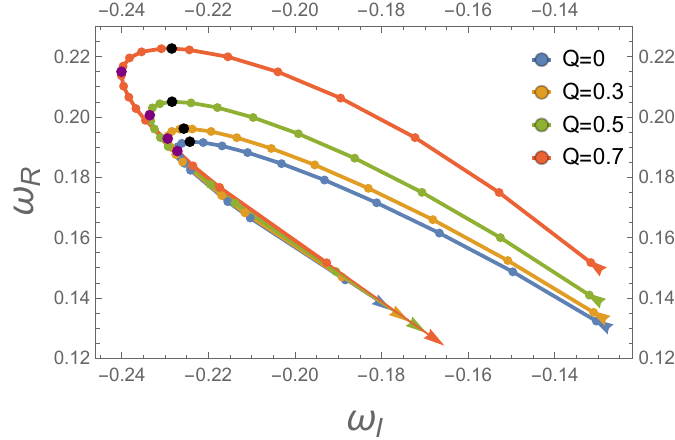}}
\subfigure[$a=0.5$]
{\label{figeps12} %% label for first subfigure
\includegraphics[width=1.7in]{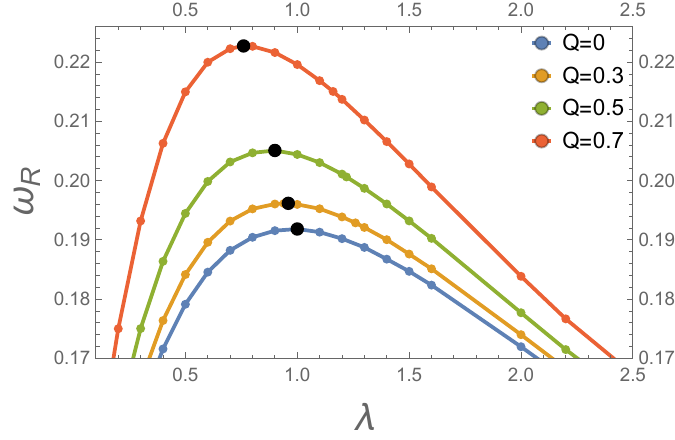}}
\subfigure[$a=0.5$]
{\label{figeps12} %% label for first subfigure
\includegraphics[width=1.83in]{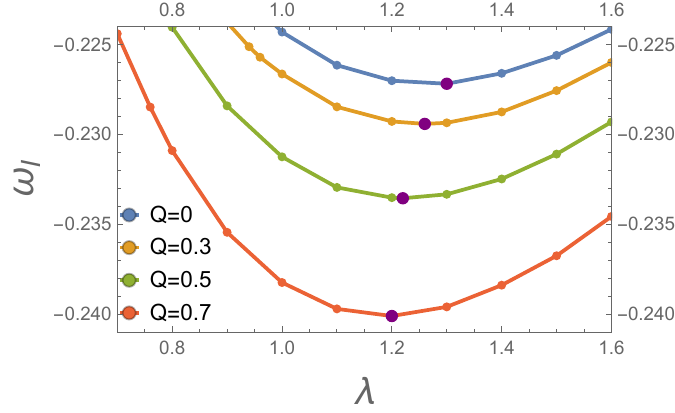}}
\subfigure[$a=-0.5$]
{\label{figeps11} %% label for first subfigure
\includegraphics[width=1.7in]{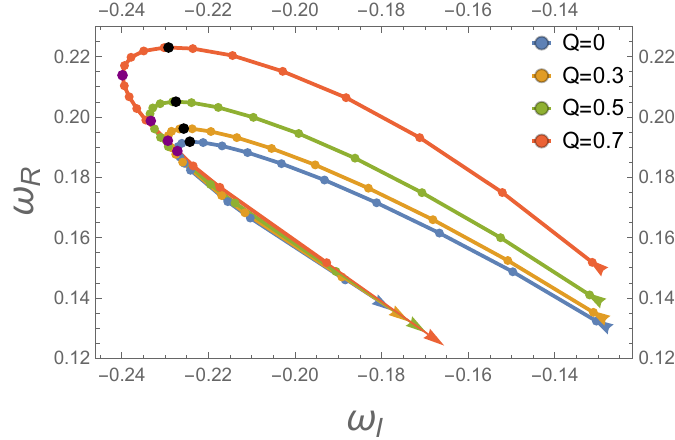}}
\subfigure[$a=-0.5$]
{\label{figeps12} %% label for first subfigure
\includegraphics[width=1.7in]{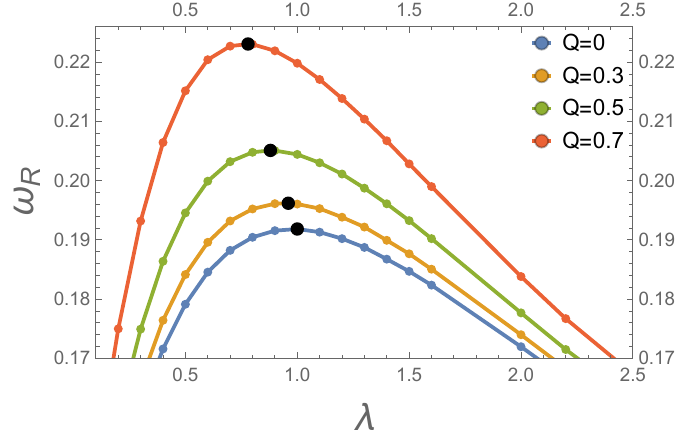}}
\subfigure[$a=-0.5$]
{\label{figeps12} %% label for first subfigure
\includegraphics[width=1.83in]{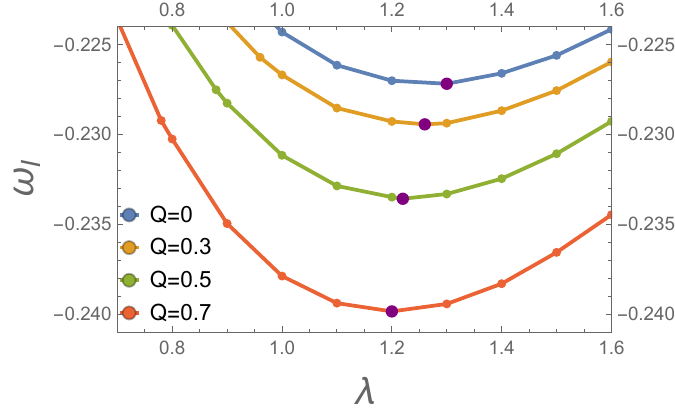}}
\caption{Variation of scalar fundamental QNFs ($n=0$) with respect to different $\lambda$ for $M=1$, $l=0$ . The arrows in panels (a) and (d) indicate the increasing direction of $\lambda$ with \(\lambda\in[0.1,3]\). Black dots denote the maxima of \(\omega_R\) and purple dots denote the minima of \(\omega_I\) versus \(\lambda\). Panels (b)--(c) and (e)--(f) show the decomposed real and imaginary parts $\omega_\text{R}$ and $\omega_\text{I}$ corresponding to panels (a) and (d), respectively, from which the values of $\lambda$ at the black and purple dots can be directly extracted.
}\label{figeps1}
\end{figure}

\begin{figure}[H]
\centering
\subfigure[$a=0.5$]
{\label{figeps11} %% label for first subfigure
\includegraphics[width=1.7in]{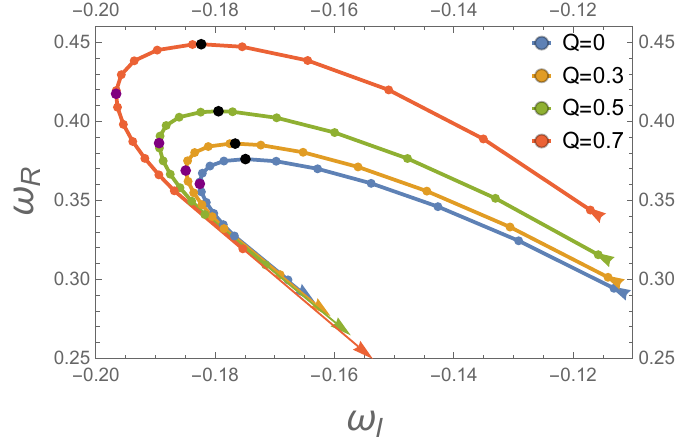}}
\subfigure[$a=0.5$]
{\label{figeps12} %% label for first subfigure
\includegraphics[width=1.7in]{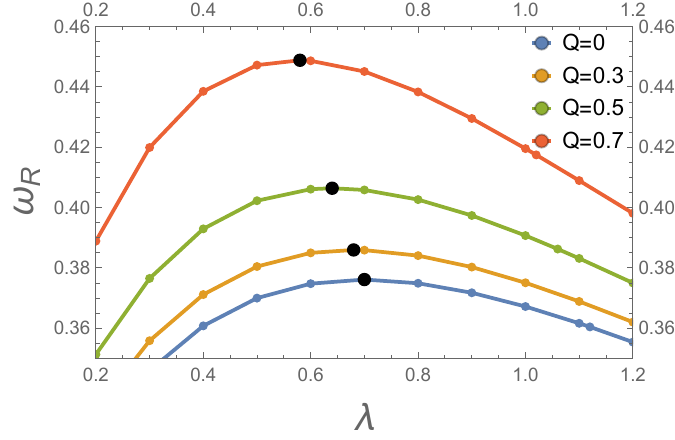}}
\subfigure[$a=0.5$]
{\label{figeps12} %% label for first subfigure
\includegraphics[width=1.83in]{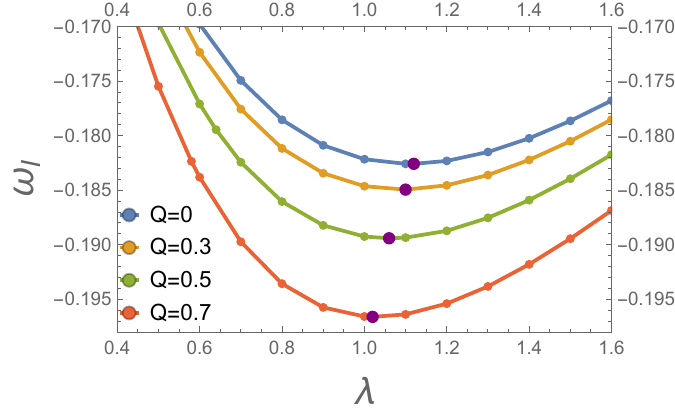}}
\subfigure[$a=-0.5$]
{\label{figeps11} %% label for first subfigure
\includegraphics[width=1.7in]{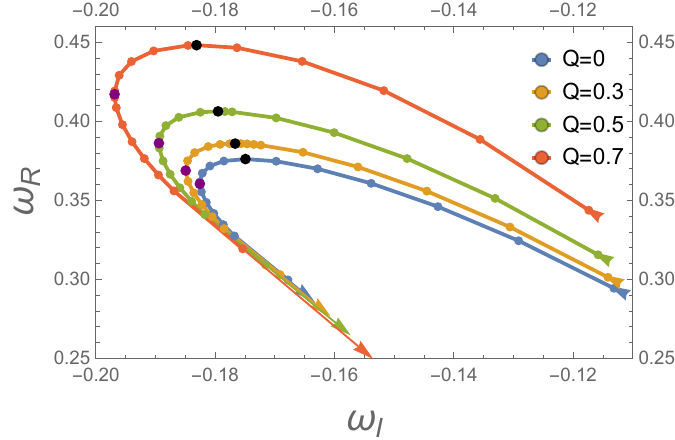}}
\subfigure[$a=-0.5$]
{\label{figeps12} %% label for first subfigure
\includegraphics[width=1.7in]{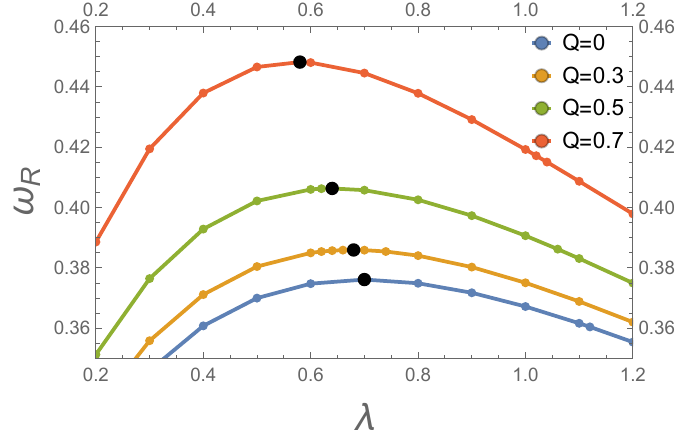}}
\subfigure[$a=-0.5$]
{\label{figeps12} %% label for first subfigure
\includegraphics[width=1.83in]{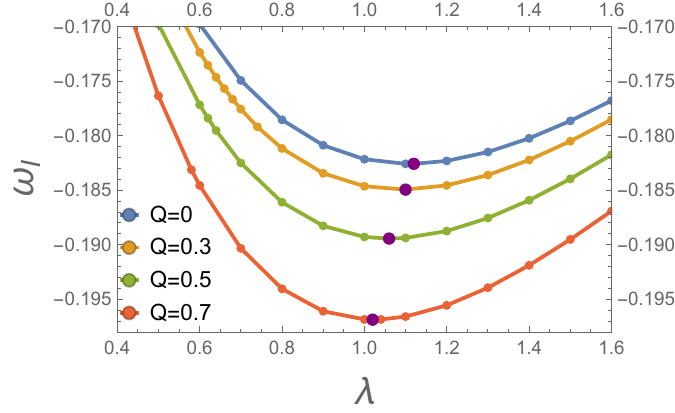}}
\caption{Variation of electromagnetic fundamental QNFs ($n=0$) with respect to different $\lambda$ for $M=1$, $l=1$ .The arrows in panels (a) and (d) indicate the increasing direction of $\lambda$ with \(\lambda\in[0.1,3]\). Black dots denote the maxima of \(\omega_R\) and purple dots denote the minima of \(\omega_I\) versus \(\lambda\). Panels (b)--(c) and (e)--(f) show the decomposed real and imaginary parts $\omega_\text{R}$ and $\omega_\text{I}$ corresponding to panels (a) and (d), respectively, from which the values of $\lambda$ at the black and purple dots can be directly extracted.
}\label{figeps1}
\end{figure}

\begin{figure}[H]
\centering
\subfigure[scalar,$l=0$]
{\label{figeps11} %% label for first subfigure
\includegraphics[width=2.8in]{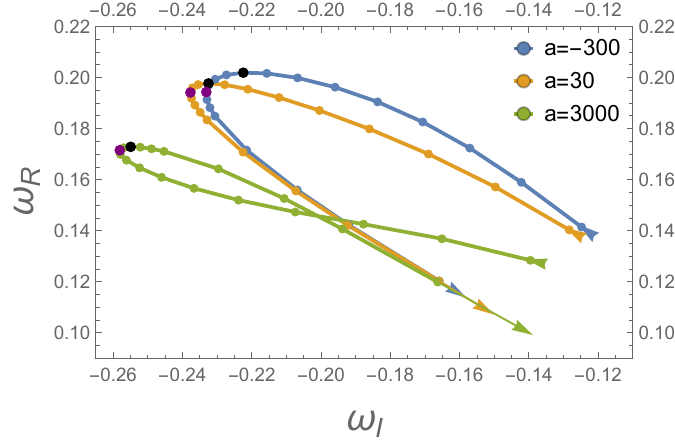}}
\subfigure[electromagnetic,$l=1$]
{\label{figeps12} %% label for first subfigure
\includegraphics[width=2.8in]{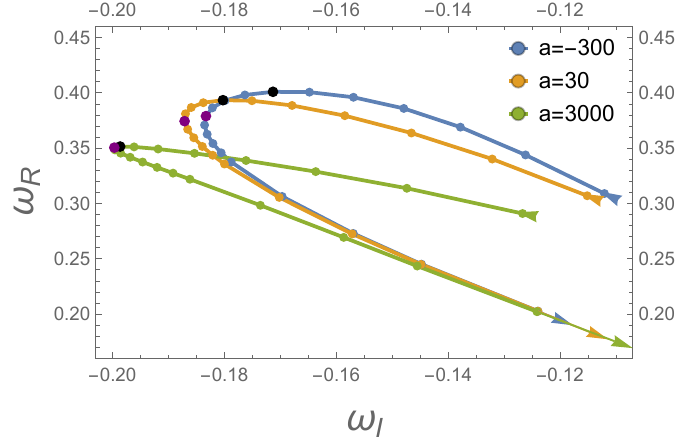}}
\caption{Variation of scalar and electromagnetic fundamental QNFs ($n=0$) with respect to different $\lambda$ for $M=1$.The arrows in panels (a) and (b) show the increasing direction of \(\lambda\) with \(\lambda\in[0.1,4]\). Black dots denote the maxima of \(\omega_R\) and purple dots denote the minima of \(\omega_I\) versus \(\lambda\). For \(a=-300,30,3000\), black dots are near \(\lambda=0.86,0.92,1.27\) and purple dots near \(\lambda=1.3,1.21,1.1\) in (a), and black dots near \(\lambda=0.67,0.73,0.8\) and purple dots near \(\lambda=1.1,1.03,0.9\) in (b).
}\label{figeps1}
\end{figure}

\subsection{\textbf{\texorpdfstring{Dependence on the nonlinear parameter $a$}{a--dependence}}}

For massless scalar and electromagnetic perturbations around Euler--Heisenberg black holes surrounded by perfect fluid dark matter with $M=1$, the fundamental quasinormal frequencies ($n=0$) show stable monotonic variation with the nonlinear parameter $a$, as shown in Figs.~8--10, with arrows pointing to the direction of increasing $a$. The real oscillation frequency $\omega_R$ of scalar perturbations decreases monotonically with rising $a$, and the imaginary damping frequency $\omega_I$ changes monotonically synchronously; electromagnetic perturbations follow the same trend but are less sensitive to $a$ than scalar ones, and the regulatory effect of $a$ on the frequencies of both perturbations is significantly suppressed as $l$ increases. Compared with the vacuum case $\lambda=0$, the dark matter environment $\lambda=0.1$ significantly amplifies the sensitivity of quasinormal frequencies to the variation of $a$, and the increase of black hole charge $Q$ further strengthens the effect of $a$, making the monotonic variation of $\omega_R$ and $\omega_I$ with $a$ more pronounced. There is a strong coupling between $a$ and $\lambda$: the increase of $\lambda$ raises the overall values of $\omega_R$ and $|\omega_I|$, the frequency difference between the first and last points in the $a$-sequence first increases and then decreases reaching the extremum at a critical $\lambda$, and in the large-$\lambda$ strong dark matter regime, the effect of $a$ is nearly negligible and the perturbation dynamics is completely dominated by the dark matter background. On the whole, $a$ only produces observable effects in the strong nonlinear regime, mainly correcting the near-horizon geometry with little influence on the global profile of the effective potential barrier.

\begin{figure}[H]
\centering
\subfigure[scalar,$\lambda=0.1$]
{\label{figeps13} %% label for first subfigure
\includegraphics[width=1.53in]{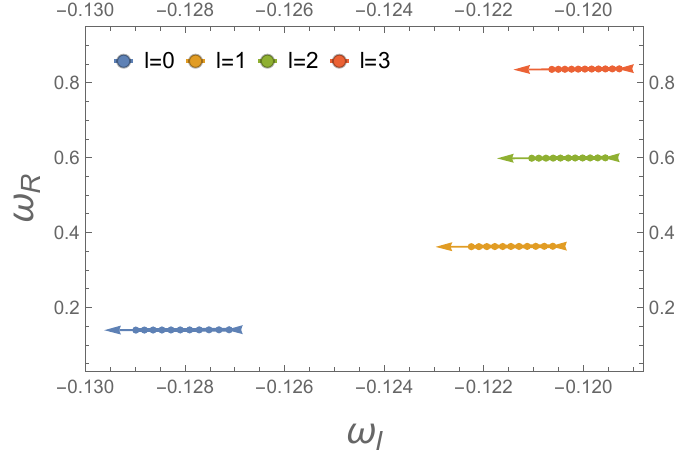}}
\subfigure[scalar,$\lambda=0$]
{\label{figeps14} %% label for first subfigure
\includegraphics[width=1.53in]{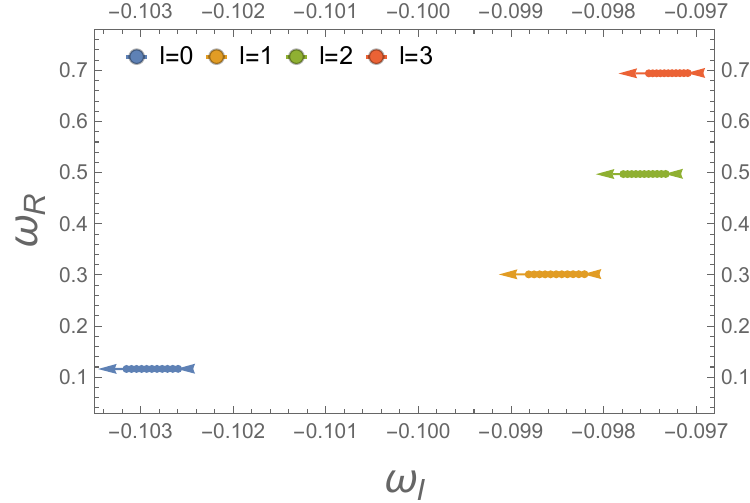}}
\subfigure[electromagnetic,$\lambda=0.1$]
{\label{figeps13} %% label for first subfigure
\includegraphics[width=1.53in]{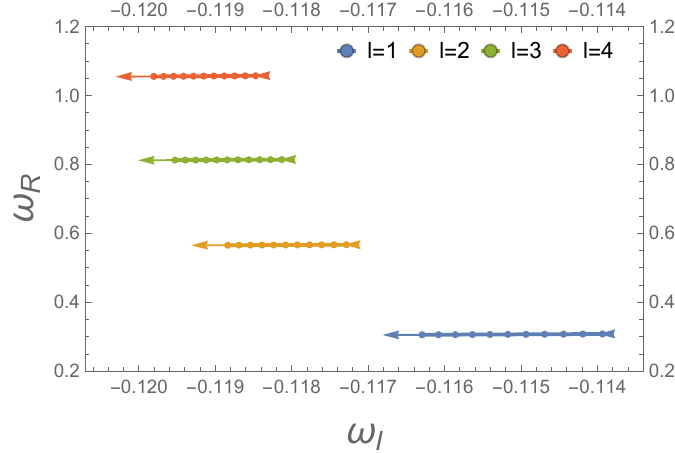}}
\subfigure[electromagnetic,$\lambda=0$]
{\label{figeps14} %% label for first subfigure
\includegraphics[width=1.53in]{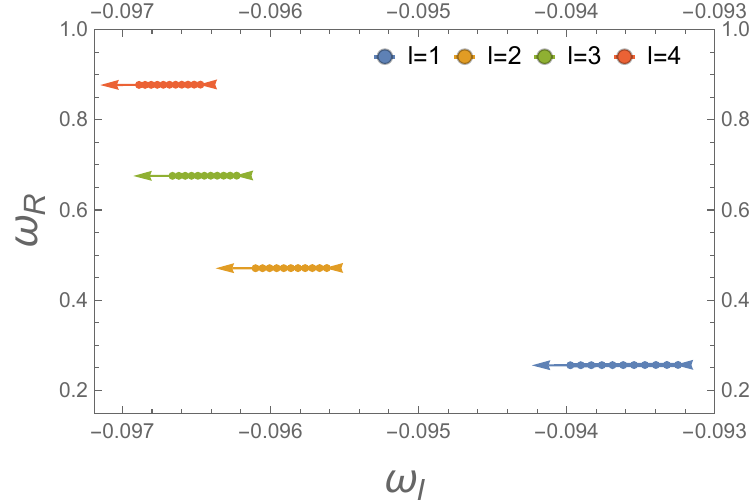}}
\caption{Variation of scalar and electromagnetic fundamental QNFs ($n=0$) with respect to different parameter $a$ for $M=1$ and $Q=0.4$. Panels (a) and (b) correspond to scalar QNFs, while panels (c) and (d) represent electromagnetic QNFs. The direction of the arrows indicates the increasing direction of parameter $a$. Curves of different colors for distinct $l$ values share identical sampling nodes of $a$, with the parameter sequence $a=-100, -80, -60, -40, -20, 0, 20, 40, 60, 80, 100$.}\label{figeps1}
\end{figure}

\begin{figure}[H]
\centering
\subfigure[scalar,$\lambda=0.1$]
{\label{figeps13} %% label for first subfigure
\includegraphics[width=1.53in]{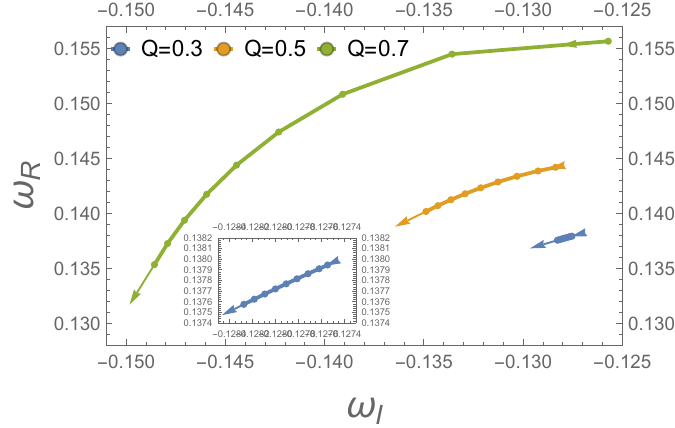}}
\subfigure[scalar,$\lambda=0$]
{\label{figeps14} %% label for first subfigure
\includegraphics[width=1.53in]{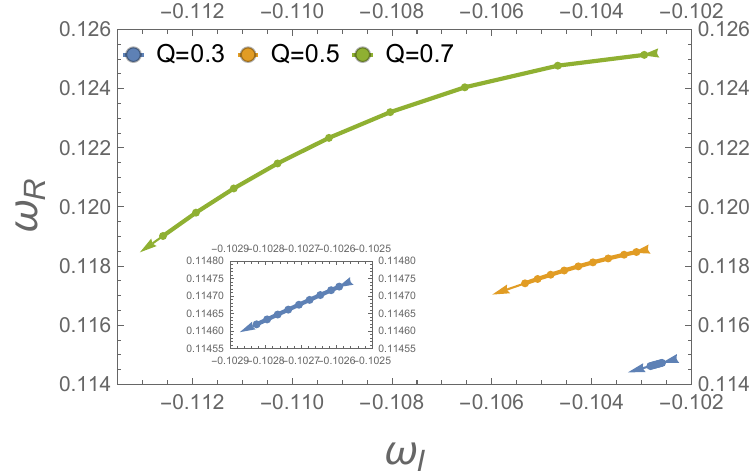}}
\subfigure[electromagnetic,$\lambda=0.1$]
{\label{figeps13} %% label for first subfigure
\includegraphics[width=1.53in]{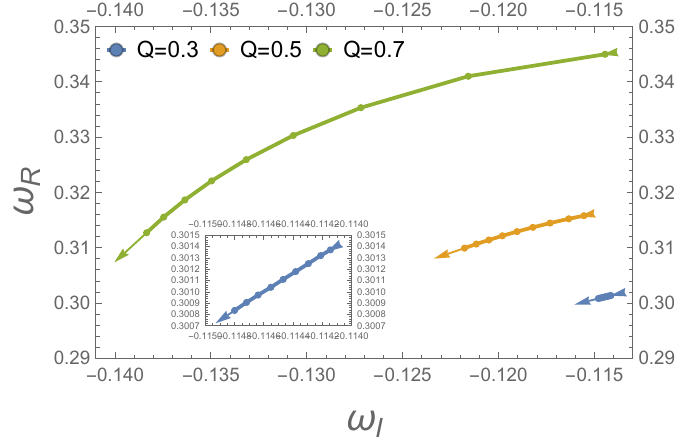}}
\subfigure[electromagnetic,$\lambda=0$]
{\label{figeps14} %% label for first subfigure
\includegraphics[width=1.53in]{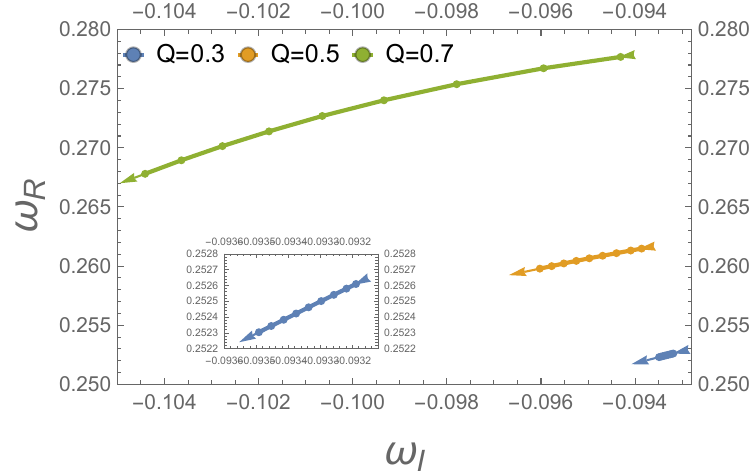}}
\caption{Variation of scalar($l=0$) and electromagnetic( $l=1$) fundamental QNFs($n=0$) with respect to different parameter $a$ for $M=1$. The direction of the arrows indicates the increasing direction of parameter $a$. Curves of different colors for distinct $Q$ values share identical sampling nodes of $a$, with the parameter sequence $a=-10, 20, 60, 100, 140, 180, 220, 260, 300$.}\label{figeps1}
\end{figure}

\begin{figure}[H]
\centering
\subfigure[scalar,$l=0$]
{\label{figeps13} %% label for first subfigure
\includegraphics[width=2.8in]{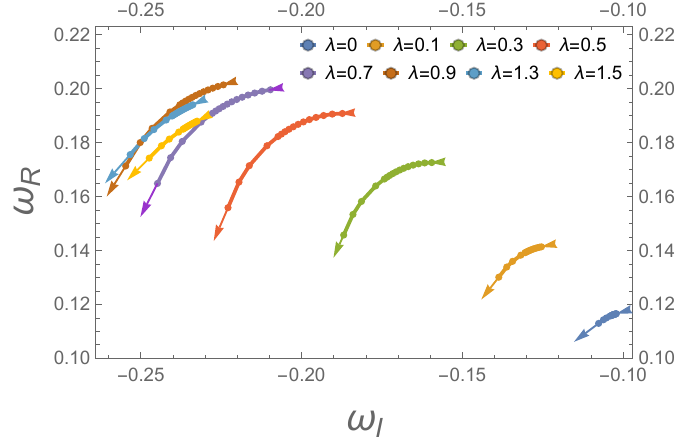}}
\subfigure[electromagnetic,$l=1$]
{\label{figeps14} %% label for first subfigure
\includegraphics[width=2.8in]{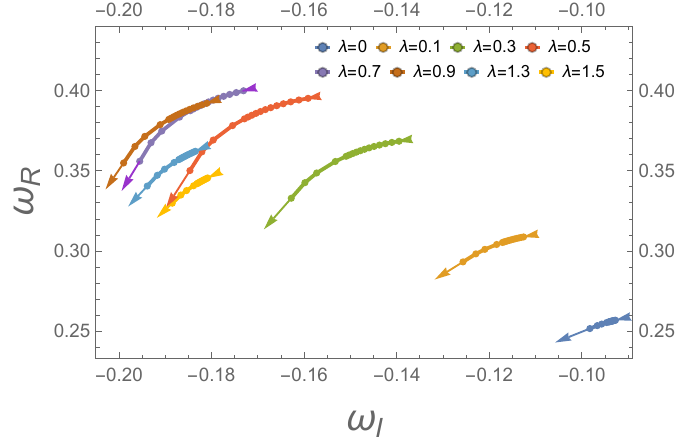}}
\caption{Variation of fundamental QNFs ($n=0$) for external fields with respect to $a$ for $M=1$ and $Q=0.4$. The arrows in panels (a) and (b) indicate the increasing direction of the parameter $a$. The sampling nodes of the curves with different colors corresponding to different $\lambda$ values are taken from the same $a$ sequence, with parameter values $a=-250, -200, -150, -100, -50, 0, 50, 100, 150, 200, 250, 300, 500, 1000, 1500, 2500$.}\label{figeps1}
\end{figure}

\subsection{\textbf{\texorpdfstring{Dependence on the black hole charge $Q$}{Q--dependence}}}

The fundamental quasinormal frequencies ($n=0$) of massless scalar and electromagnetic perturbations around Euler--Heisenberg black holes surrounded by perfect fluid dark matter with $M=1$ present the coupled feature of monotonic oscillation frequency and nonmonotonic damping rate with the black hole charge $Q$, as depicted in Figs.~11--14, where arrows indicate the direction of increasing $Q$ and black dots mark the critical charge $Q$ corresponding to the minimum of $\omega_I$. The rise of $Q$ elevates the height of the effective potential barrier and strengthens the confinement of perturbation waves near the photon sphere, so the real oscillation frequency $\omega_R$ increases monotonically with $Q$, and low-multipole perturbations are more sensitive to the change of $Q$; the imaginary damping frequency $\omega_I$ decreases first to the minimum at $Q$ and then increases with $Q$, originating from the competition between the height of the effective potential barrier and the curvature near the peak. The increase of dark matter parameter $\lambda$ monotonically raises $Q_c$, and when $\lambda\gtrsim1$, the nonmonotonic damping feature disappears completely and $\omega_I$ changes monotonically with $Q$, which holds for both scalar and electromagnetic perturbations and is nearly independent of the sign of $a$. In the strong nonlinear regime, the increase of $a$ changes the dependence of $\omega_R$ on $Q$ from monotonic increase to first increase and then decrease, makes the evolution of $\omega_I$ more complex, and monotonically raises the critical extremum charge $Q_c$; scalar perturbations are more sensitive to $Q$ than electromagnetic ones due to the extra derivative term in the effective potential. Overall, $Q$ dominates the growth of oscillation frequency and regulates the damping behavior through the curvature of the effective potential barrier, while $\lambda$ and $a$ act as environmental and nonlinear factors respectively to reshape the damping structure and the position of the extremum charge.

\begin{figure}[H]
\centering
\subfigure[$a=0.5$,$\lambda=0.1$]
{\label{figeps11} %% label for first subfigure
\includegraphics[width=1.53in]{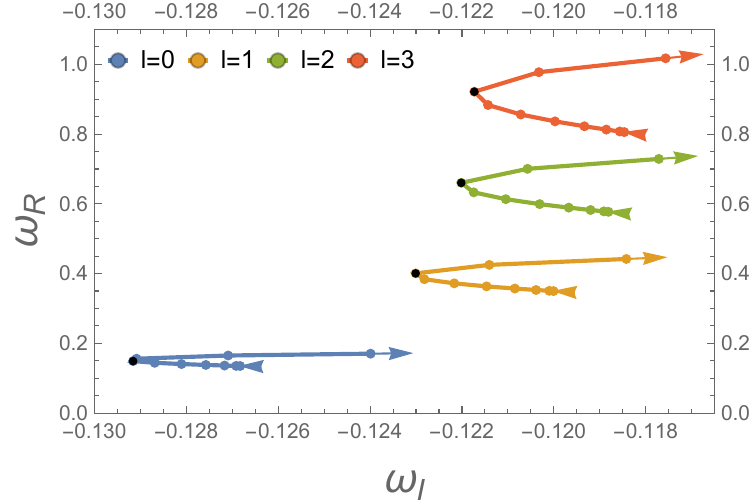}}
\subfigure[$a=-0.5$,$\lambda=0.1$]
{\label{figeps12} %% label for first subfigure
\includegraphics[width=1.53in]{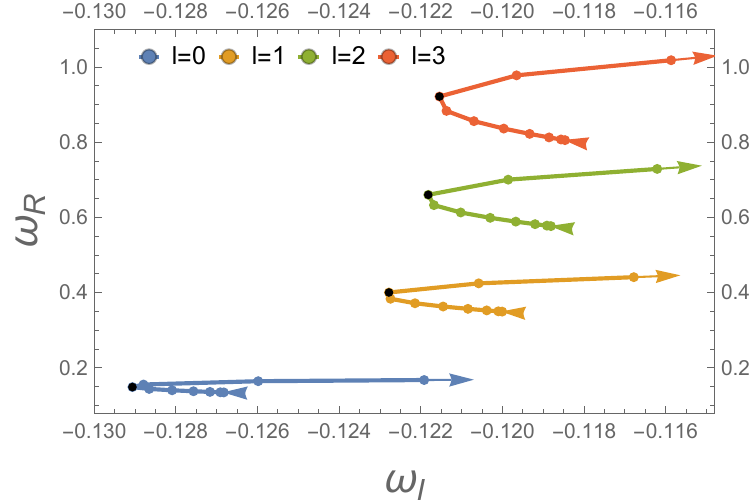}}
\subfigure[$a=0.5$,$\lambda=0$]
{\label{figeps13} %% label for first subfigure
\includegraphics[width=1.53in]{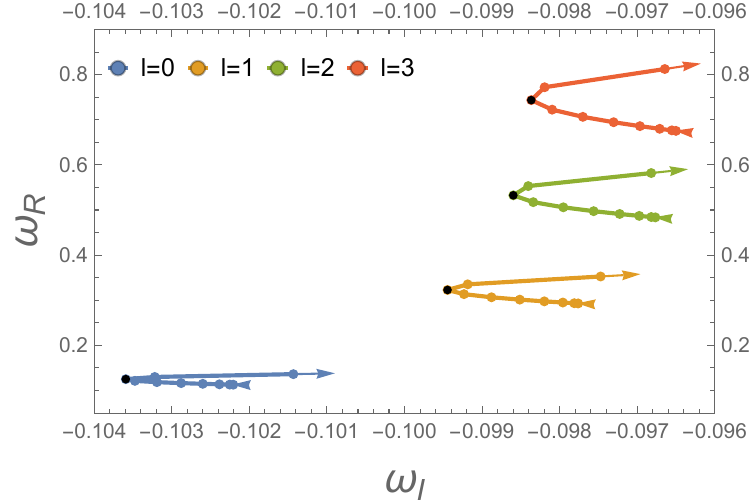}}
\subfigure[$a=-0.5$,$\lambda=0$]
{\label{figeps14} %% label for first subfigure
\includegraphics[width=1.53in]{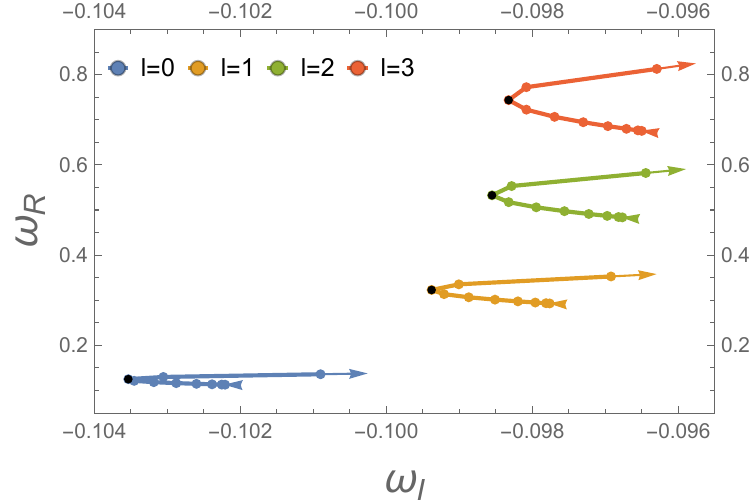}}
\caption{Variation of scalar fundamental QNFs ($n=0$) with respect to different $Q$ for $M=1$.The arrows in panels (a)–(d) show the increasing direction of \(Q\in[0,Q_c)\) (\(Q_c\) is the extreme parameter). Black dots denote the minima of \(\omega_I\) versus \(Q\). In (a)–(b), black dots are near \(Q\approx0.7\) for \(l=1,2,3\) and \(Q\approx0.6\) for \(l=0\). In (c)–(d), black dots are near \(Q\approx0.7\) for \(l=0,1,2,3\).}\label{figeps1}
\end{figure}

\begin{figure}[H]
\centering
\subfigure[$a=0.5$,$\lambda=0.1$]
{\label{figeps11} %% label for first subfigure
\includegraphics[width=1.53in]{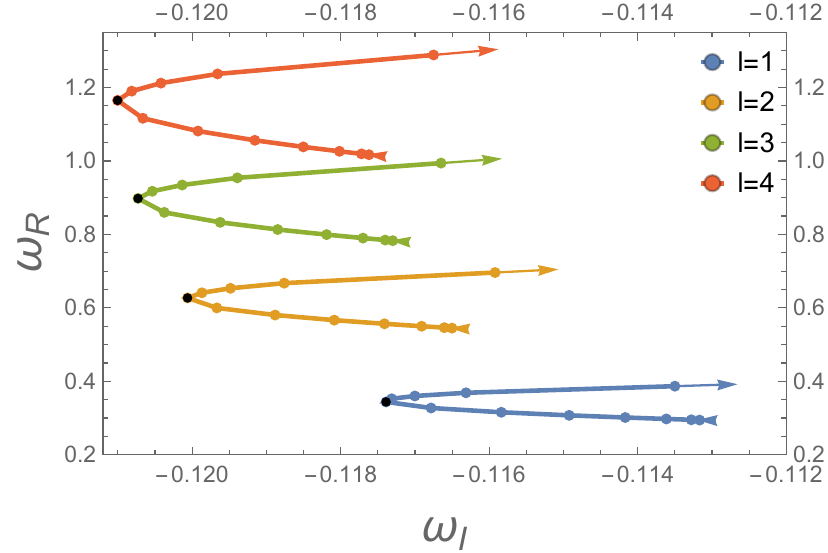}}
\subfigure[$a=-0.5$$\lambda=0.1$]
{\label{figeps12} %% label for first subfigure
\includegraphics[width=1.53in]{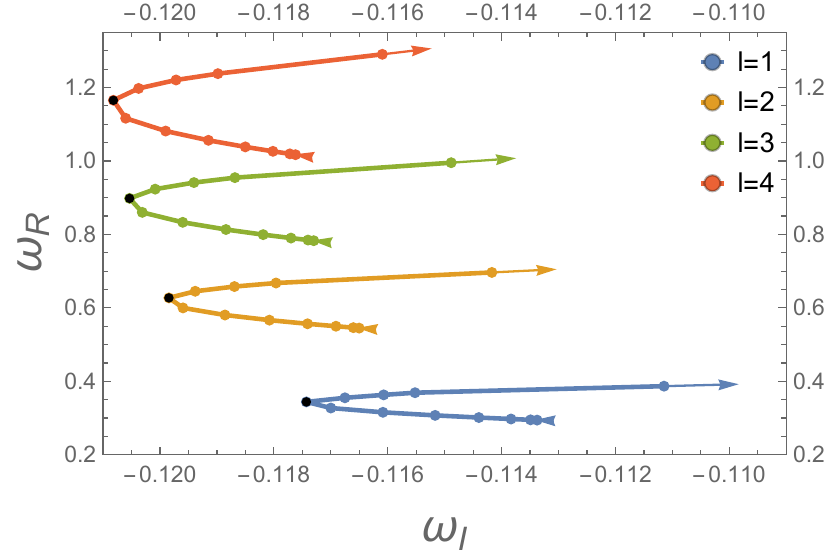}}
\subfigure[$a=0.5$,$\lambda=0$]
{\label{figeps13} %% label for first subfigure
\includegraphics[width=1.53in]{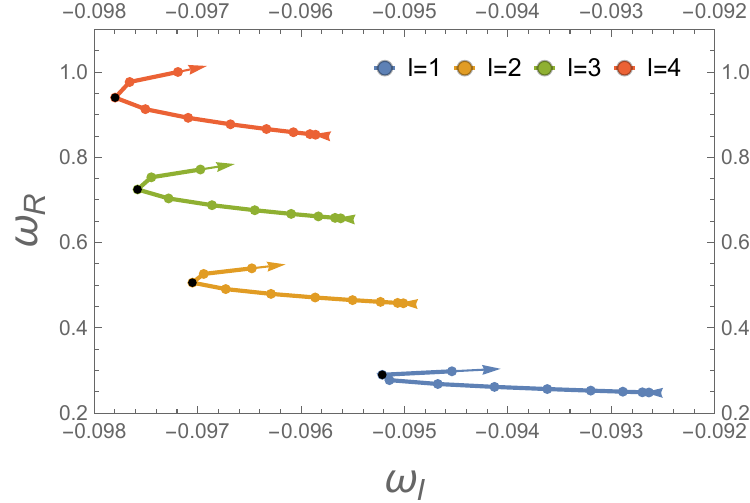}}
\subfigure[$a=-0.5$,$\lambda=0$]
{\label{figeps14} %% label for first subfigure
\includegraphics[width=1.53in]{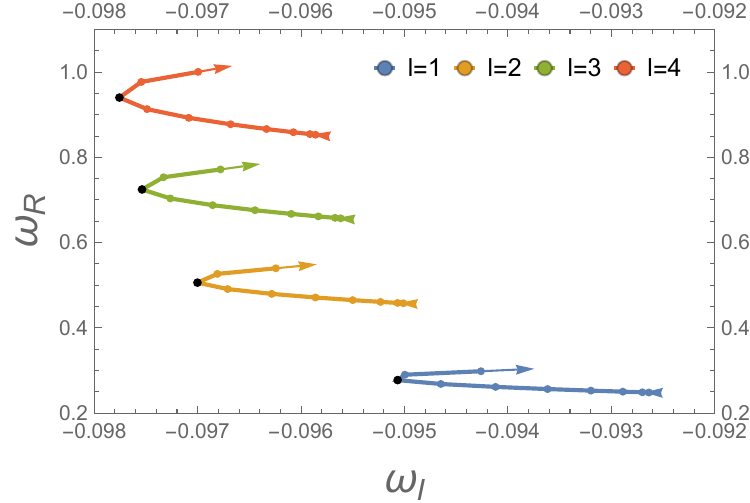}}
\caption{Variation of electromagnetic fundamental QNFs ($n=0$) with respect to different $Q$ for $M=1$.The arrows in panels (a)–(d) show the increasing direction of \(Q\in[0,Q_c)\) (\(Q_c\) is the extreme parameter). Black dots denote the minima of \(\omega_I\) versus \(Q\). In (a)–(b), black dots are near \(Q\approx0.7\) for \(l=0,1,2,3\). In (c)–(d), black dots are near \(Q\approx0.7\) for \(l=2,3,4\) and \(Q\approx0.8\) for \(l=1\).
}\label{figeps1}
\end{figure}

\begin{figure}[H]
\centering
\subfigure[$a=0.5$,$l=0$]
{\label{figeps11} %% label for first subfigure
\includegraphics[width=1.53in]{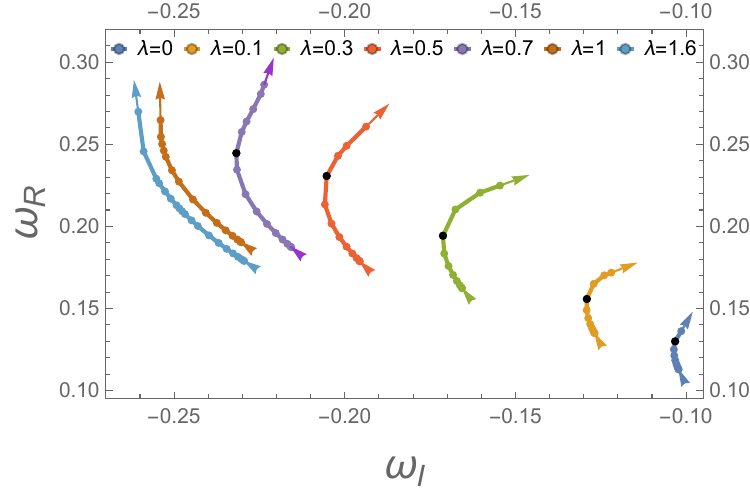}}
\subfigure[$a=-0.5$,$l=0$]
{\label{figeps12} %% label for first subfigure
\includegraphics[width=1.53in]{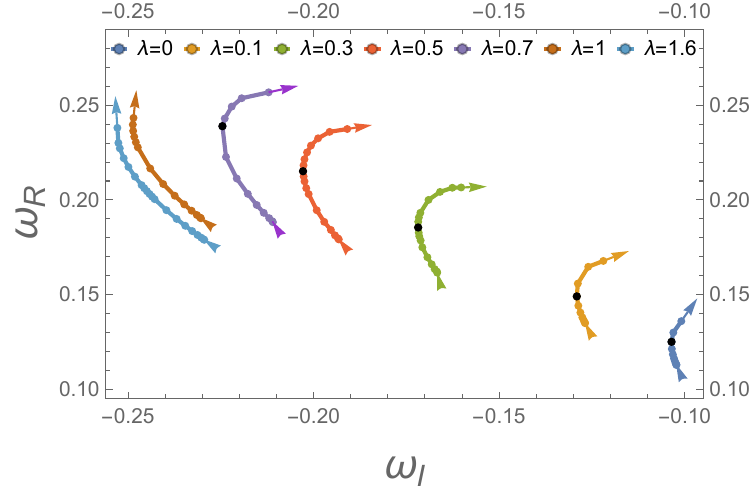}}
\subfigure[$a=0.5$,$l=1$]
{\label{figeps11} %% label for first subfigure
\includegraphics[width=1.53in]{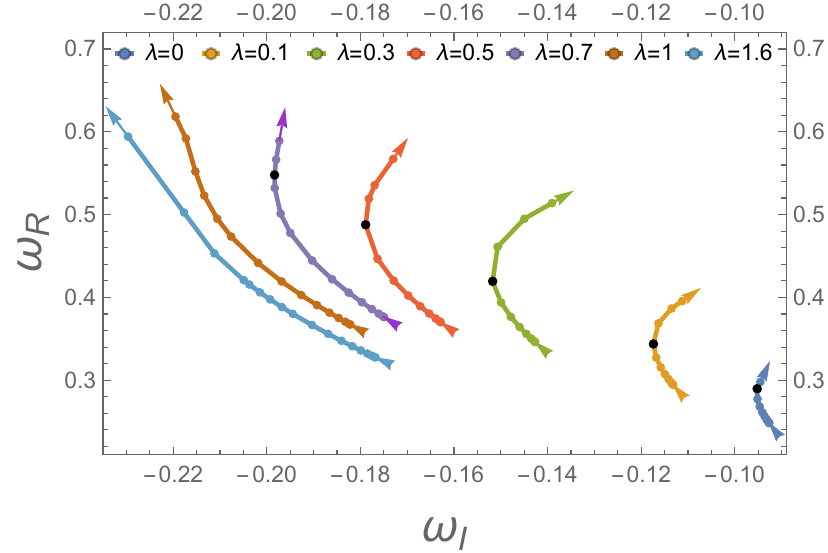}}
\subfigure[$a=-0.5$,$l=1$]
{\label{figeps12} %% label for first subfigure
\includegraphics[width=1.53in]{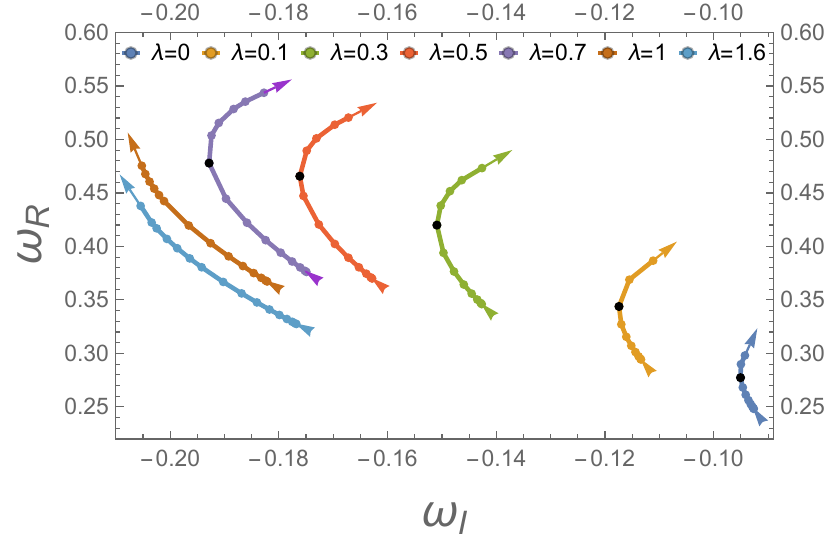}}
\caption{Variation of scalar ($l=0$) and electromagnetic ($l=1$) fundamental QNFs ($n=0$) with respect to different $Q$ for $M=1$. Arrows indicate the increasing direction of $Q\in[0,Q_c)$ ($Q_c$ rises with $\lambda$). Black dots mark the minima of $\omega_I$ against $Q$. For scalar QNFs, minima emerge near $Q=0.7,0.6,0.68,0.72,0.85$ (panel (a)) and $Q=0.7,0.6,0.63,0.7,0.8$ (panel (b)) for $\lambda=0,0.1,0.3,0.5,0.7$, with no minima for $\lambda=1,1.6$. For electromagnetic QNFs, minima appear near $Q=0.8,0.69,0.71,0.8,0.92$ (panel (c)) and $Q=0.7,0.69,0.71,0.76,0.8$ (panel (d)) for the same $\lambda$ values, and vanish at $\lambda=1,1.6$.
}\label{figeps1}
\end{figure}

\begin{figure}[H]
\centering
\subfigure[scalar,$l=0$]
{\label{figeps13} %% label for first subfigure
\includegraphics[width=2.5in]{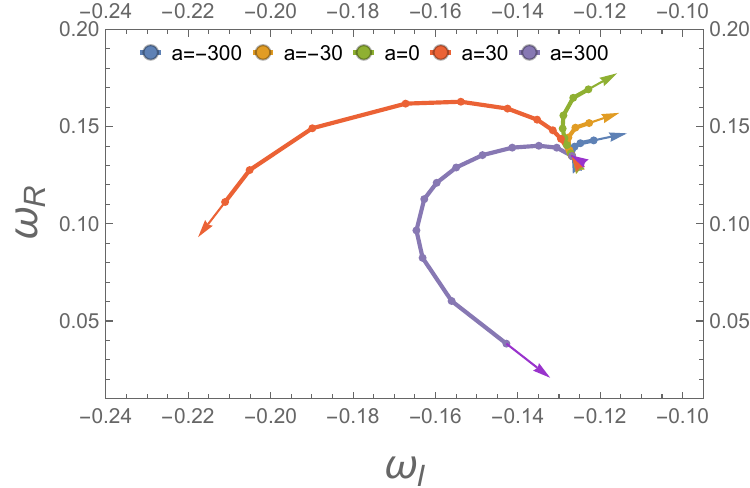}}
\subfigure[electromagnetic,$l=1$]
{\label{figeps14} %% label for first subfigure
\includegraphics[width=2.5in]{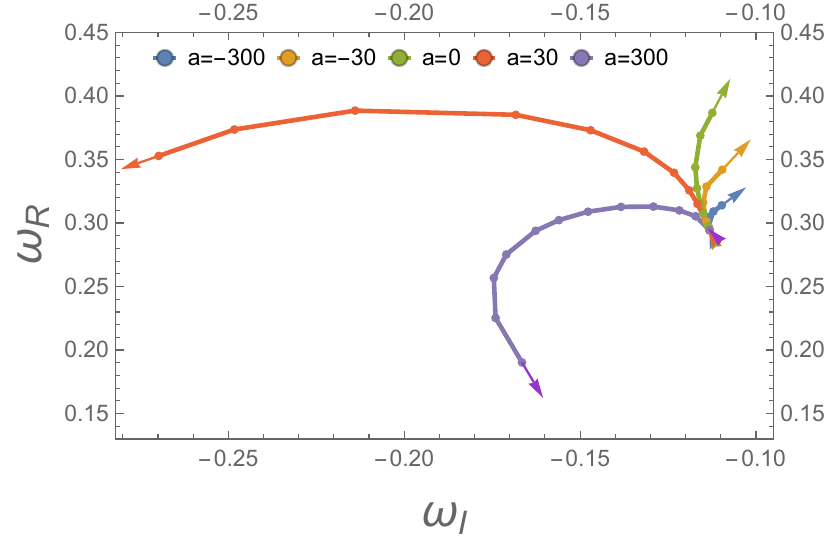}}
\caption{Variation of fundamental QNFs ($n=0$) for external fields with respect to different $Q$ for $M=1$ and $\lambda=0.1$.The arrows in panels (a) and (b) indicate the increasing direction of $Q\in[0,Q_\text{c})$, where the extremal parameter $Q_\text{c}$ rises with $a$. For the curves corresponding to $a=-300$, $-30$, $0$, $30$, and $300$, the respective ranges of $Q$ are $[0,0.45]$, $[0,0.68]$, $[0,0.85]$, $[0,1.6]$, and $[0,2.4]$.}\label{figeps1}
\end{figure}

\section{ The Greybody factor}
\label{sec4}

In this section, we use the WKB approximation method to calculate greybody factors for scalar perturbation. 
The boundary condition for the scattering process is different from that of the QNMs, which can be written as~\cite{Kokkotas:1999bd,Konoplya:2011qq}
\begin{eqnarray}
\psi& =& T(\omega)e^{-i\omega r_*}, \quad r_* \rightarrow -\infty\nonumber,\\
\psi& =& e^{-i\omega r_*} + R(\omega)e^{i\omega r_*}, \quad r_* \rightarrow +\infty,\label{boundgrey}
\end{eqnarray}
where $R$ and $T$ represent the reflection coefficient and transmission coefficient, respectively. The greybody factor is defined as the probability of an outgoing wave
reaching to infinity or an incoming wave absorbed by the black hole~\cite{Page:1976df,Harmark:2007jy,Konoplya:2021ube}. Therefore, $|T(\omega)|^2$ is called the greybody factor, and $R(\omega)$ and $T(\omega)$ should satisfy the following relation
\begin{eqnarray}
|R(\omega)|^2+|T(\omega)|^2=1.
\end{eqnarray}

Using the 6th-order WKB method, the
reflection and transmission coefficients can be obtained~\cite{Iyer:1986np,Konoplya:2003ii}
\begin{eqnarray}
&&|R(\omega)|^2 = \frac{1} {1 + e^{-2\pi i K(\omega)}} ,\nonumber\\
&&|T(\omega)|^2 =\frac{1} {1 + e^{2\pi i K(\omega)}}= 1 -|R(\omega)|^2,\label{TR}
\end{eqnarray}
where $K$ is a parameter which can be obtained by the WKB formula~\cite{Konoplya:2003ii}
\begin{eqnarray}
K= \frac{i\left( \omega^2 - V(r_0) \right)}{\sqrt{-2V''(r_0)}} + \sum_{i=2}^6 \Lambda_i.
\end{eqnarray}
Figures 15--20 present the greybody factors $|T|^{2}$ for scalar and electromagnetic perturbations in the Euler--Heisenberg black hole spacetime surrounded by perfect fluid dark matter (PFDM). In all cases, the transmission probability increases monotonically with the wave frequency $\omega$, approaching zero in the low-frequency regime and unity at high frequencies, which reflects the filtering effect of the effective potential barrier on propagating waves. The results show that the black hole charge $Q$ and the dark matter parameter $\lambda$ significantly affect the greybody spectrum by suppressing the low-frequency transmission probability and shifting the transition region toward higher frequencies, indicating a stronger scattering barrier. In contrast, the Euler--Heisenberg nonlinear parameter $a$ produces only relatively weak corrections, mainly in the low-frequency region. Moreover, for sufficiently large values of the dark matter parameter $(\lambda \gtrsim 1.4)$, the suppression effect gradually saturates and the corresponding curves become less sensitive to further increases in $\lambda$. Overall, both scalar and electromagnetic perturbations exhibit qualitatively similar behaviors, implying that the greybody factors are governed predominantly by the background spacetime geometry rather than by the spin structure of the perturbing field.

\begin{figure}[H]
\centering
\subfigure[$a=0.5,\lambda=0.1$]{\label{fig1g}%% label for first subfigure
\includegraphics[width=1.5in]{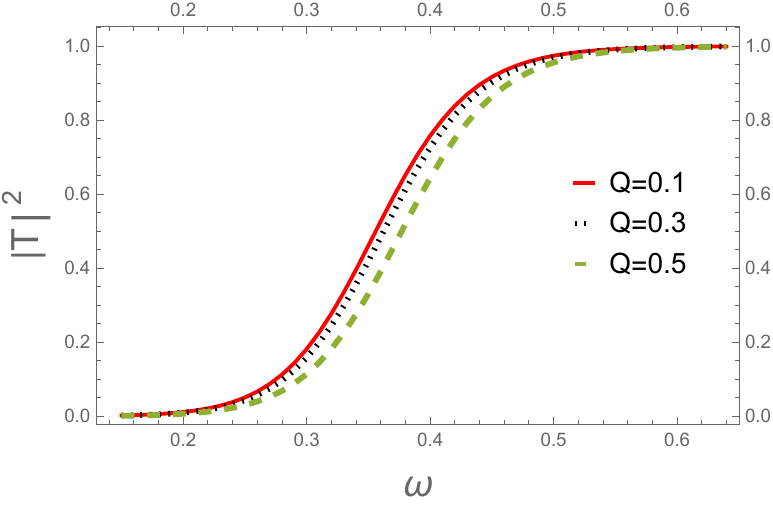}}
\subfigure[$a=-0.5,\lambda=0.1$]{\label{fig3g}%% label for first subfigure
\includegraphics[width=1.5in]{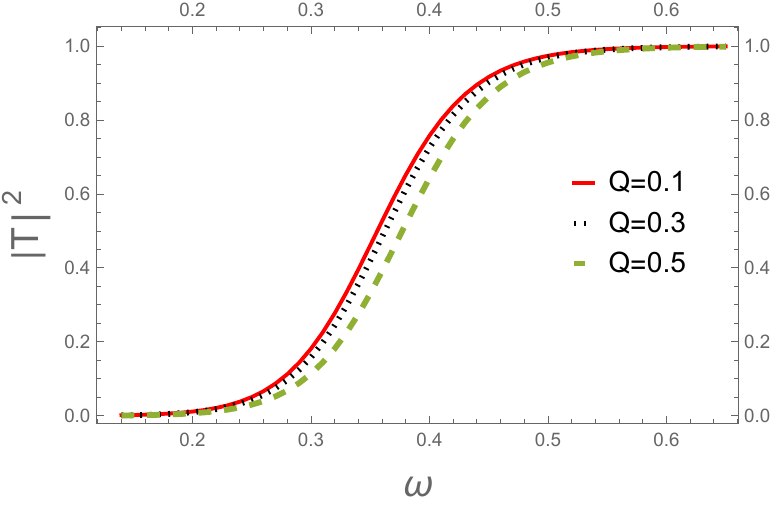}}
\subfigure[$a=0.5,\lambda=0$]{\label{fig1g}%% label for first subfigure
\includegraphics[width=1.5in]{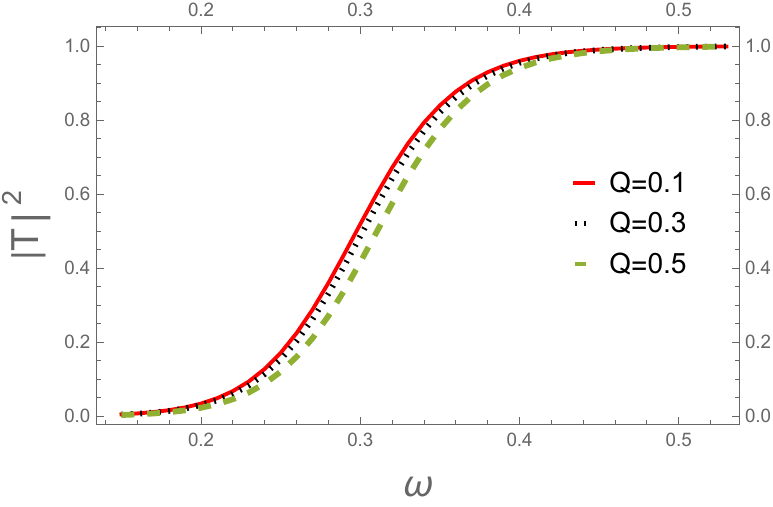}}
\subfigure[$a=-0.5,\lambda=0$]{\label{fig3g}%% label for first subfigure
\includegraphics[width=1.5in]{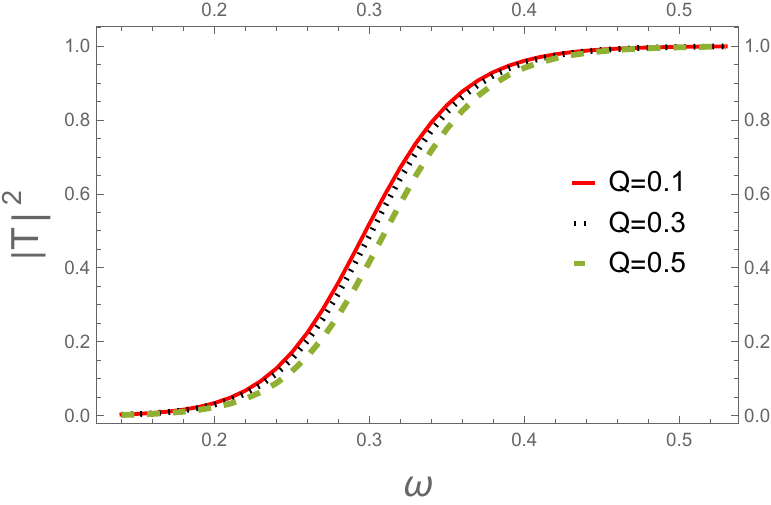}}
\caption{The scalar field greybody factor \textit{vs.} $Q$with $M=1$,$l=1$.}\label{figs1}
\end{figure}

\begin{figure}[H]
\centering
\subfigure[$a=0.5,\lambda=0.1$]{\label{fig1g}%% label for first subfigure
\includegraphics[width=1.5in]{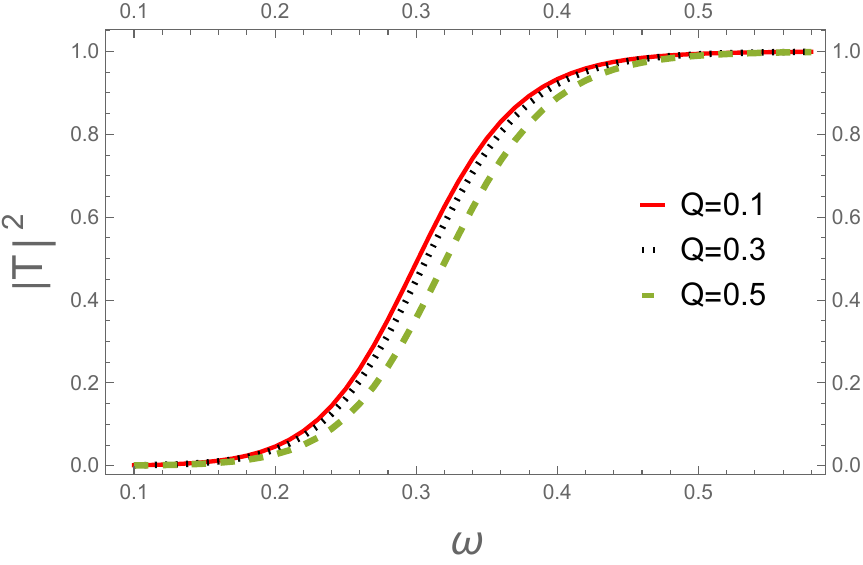}}
\subfigure[$a=-0.5,\lambda=0.1$]{\label{fig3g}%% label for first subfigure
\includegraphics[width=1.5in]{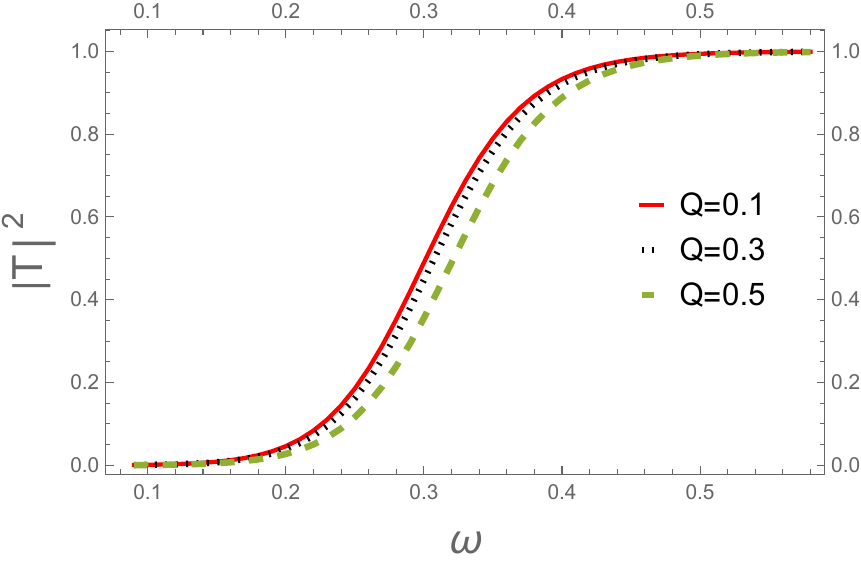}}
\subfigure[$a=0.5,\lambda=0$]{\label{fig1g}%% label for first subfigure
\includegraphics[width=1.5in]{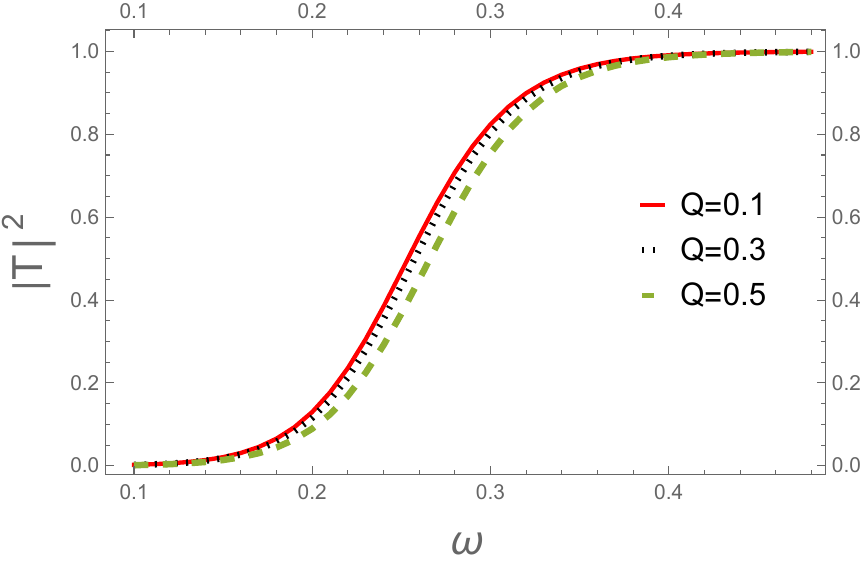}}
\subfigure[$a=-0.5,\lambda=0$]{\label{fig3g}%% label for first subfigure
\includegraphics[width=1.5in]{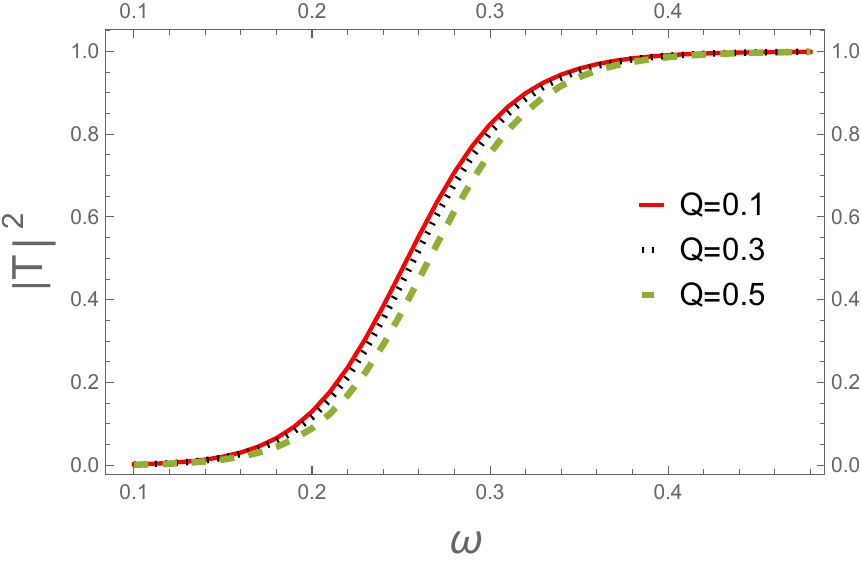}}
\caption{The electromagnetic field greybody factor \textit{vs.} $Q$ with $M=1$,$l=1$.}\label{figs1}
\end{figure}

\begin{figure}[H]
\centering
\subfigure[$\lambda=0.1$,$Q=0.4$]{\label{fig1g}%% label for first subfigure
\includegraphics[width=1.5in]{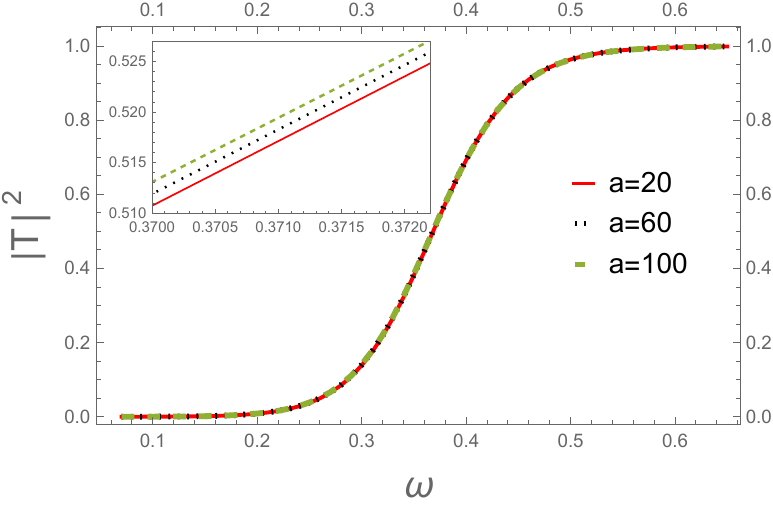}}
\subfigure[$\lambda=0.1$,$Q=0.4$]{\label{fig2g} %% label for first subfigure
\includegraphics[width=1.5in]{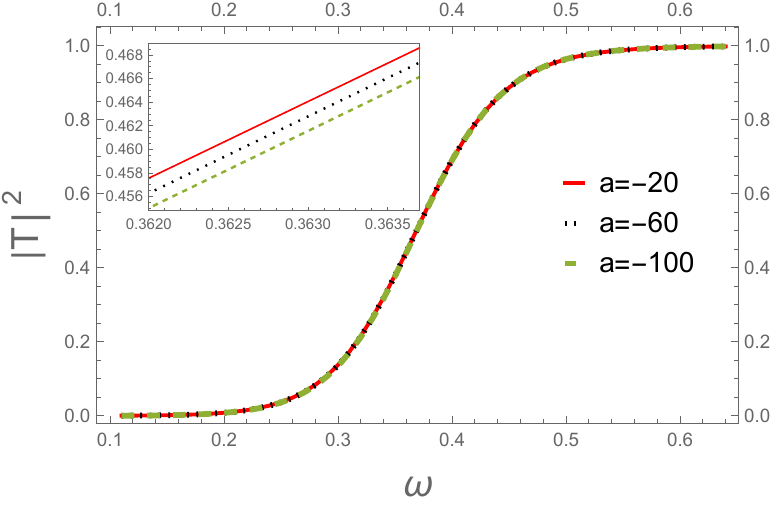}}
\subfigure[$\lambda=0$,$Q=0.4$]{\label{fig1g}%% label for first subfigure
\includegraphics[width=1.5in]{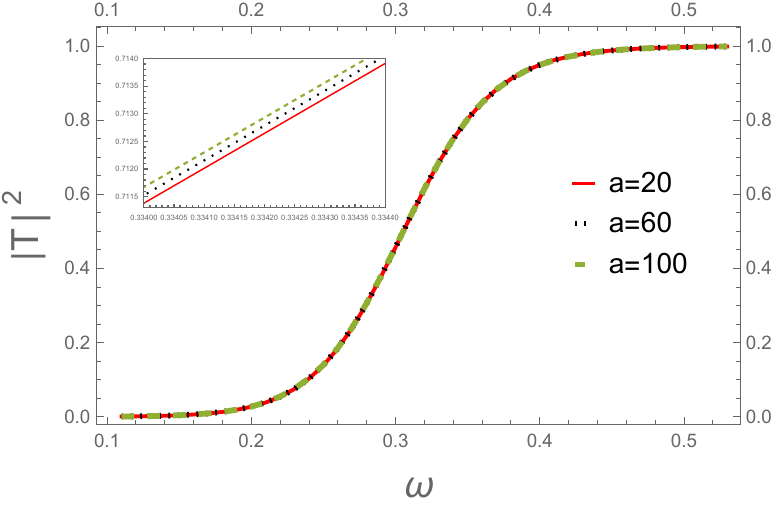}}
\subfigure[$\lambda=0$,$Q=0.4$]{\label{fig2g} %% label for first subfigure
\includegraphics[width=1.5in]{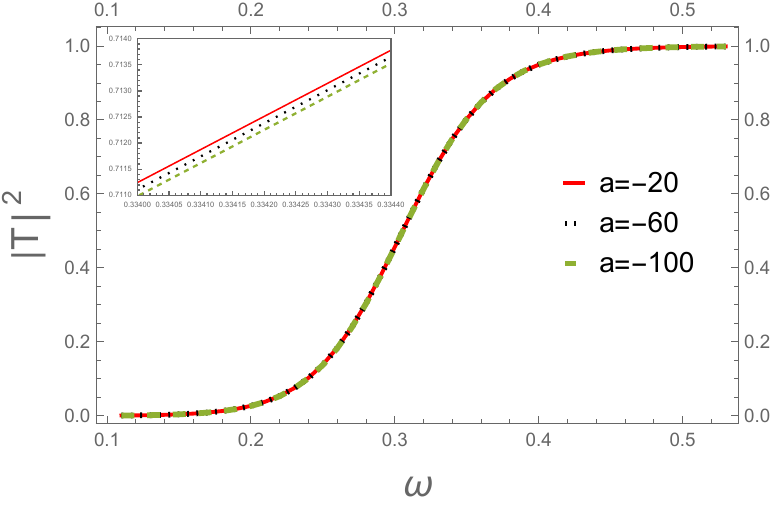}}
\caption{The scalar field greybody factor \textit{vs.} $a$ with $M=1$,$l=1$.}\label{figs1}
\end{figure}

\begin{figure}[H]
\centering
\subfigure[$\lambda=0.1$,$Q=0.4$]{\label{fig1g}%% label for first subfigure
\includegraphics[width=1.5in]{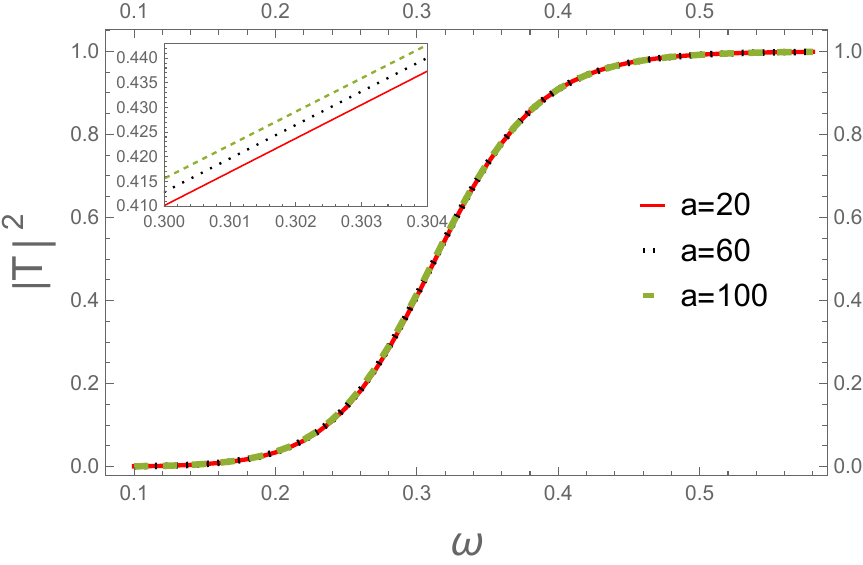}}
\subfigure[$\lambda=0.1$,$Q=0.4$]{\label{fig2g} %% label for first subfigure
\includegraphics[width=1.5in]{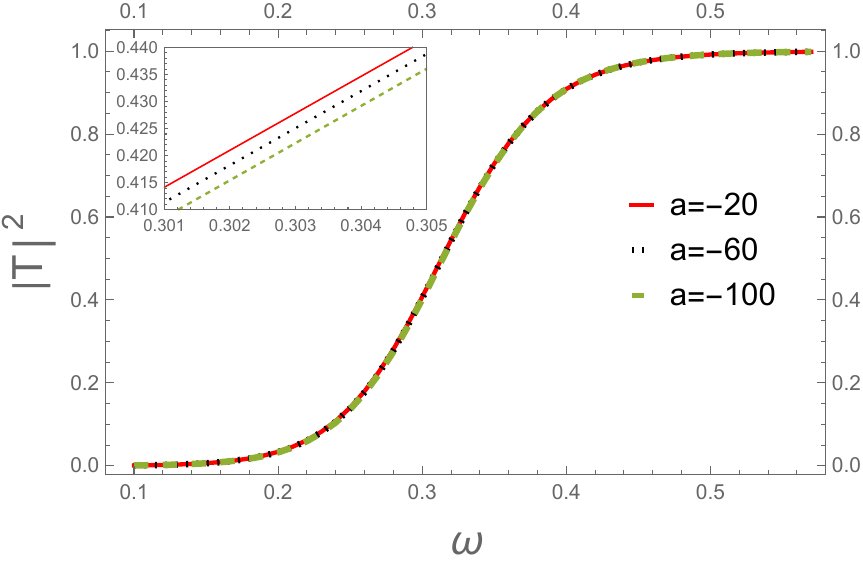}}
\subfigure[$\lambda=0$,$Q=0.4$]{\label{fig1g}%% label for first subfigure
\includegraphics[width=1.5in]{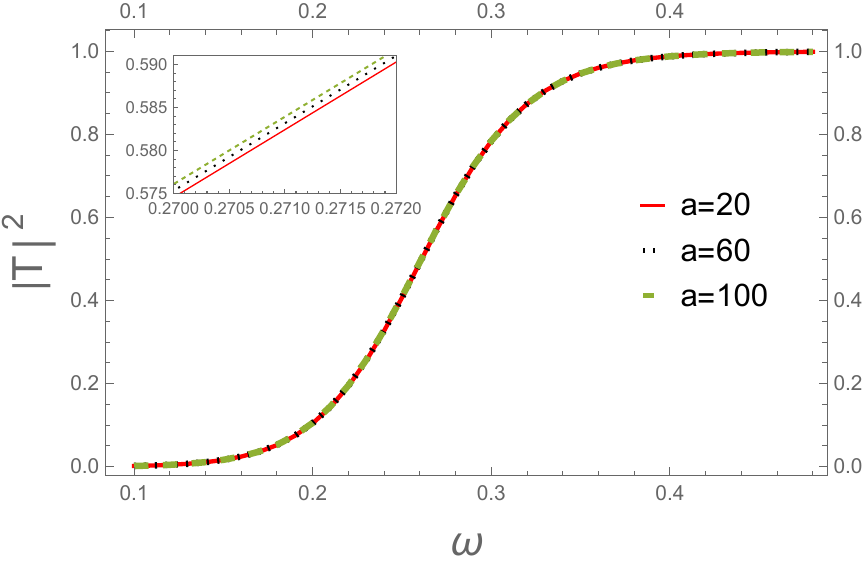}}
\subfigure[$\lambda=0$,$Q=0.4$]{\label{fig2g} %% label for first subfigure
\includegraphics[width=1.5in]{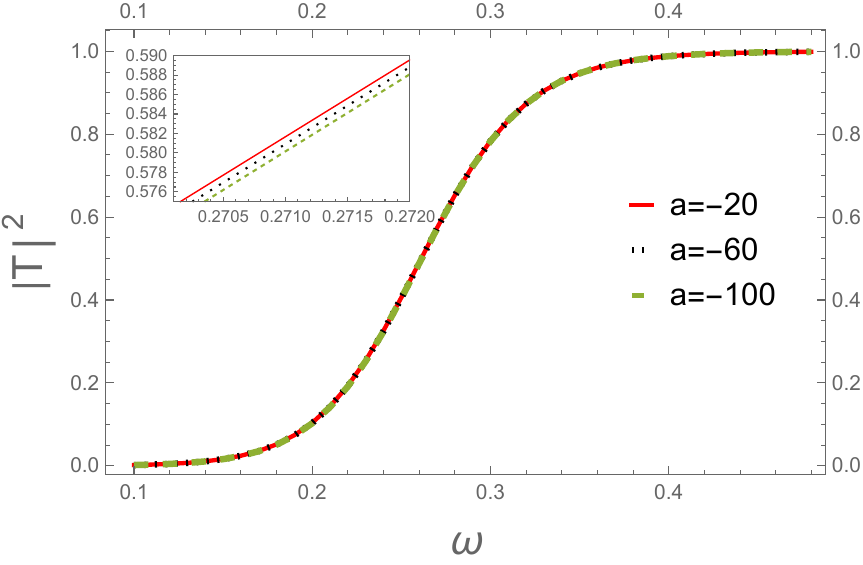}}
\caption{The electromagnetic field greybody factor \textit{vs.} $a$ with $M=1$,$l=1$.}\label{figs1}
\end{figure}

\begin{figure}[H]
\centering
\subfigure[$a=0.5$,$Q=0.4$]{\label{fig1g}%% label for first subfigure
\includegraphics[width=1.5in]{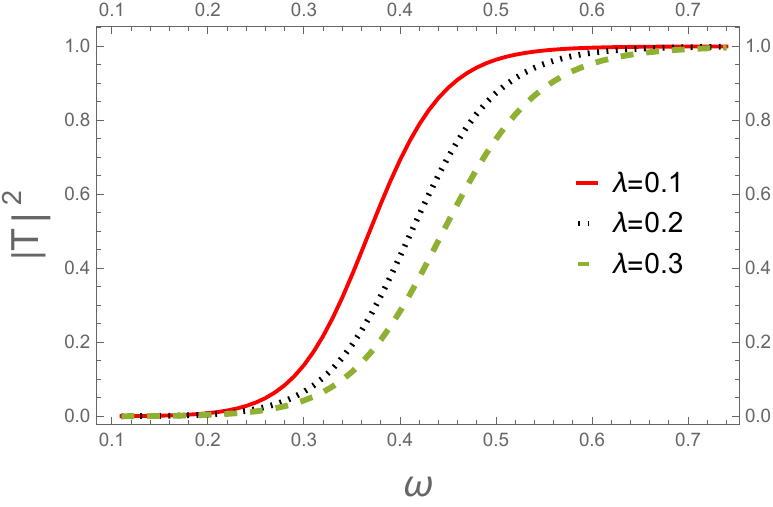}}
\subfigure[$a=-0.5$,$Q=0.4$]{\label{fig3g}%% label for first subfigure
\includegraphics[width=1.5in]{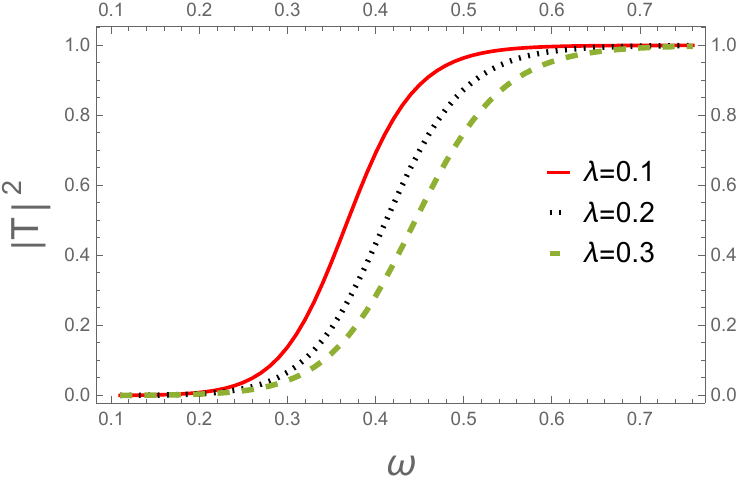}}
\subfigure[$a=0.5$,$Q=0.4$]{\label{fig1g}%% label for first subfigure
\includegraphics[width=1.5in]{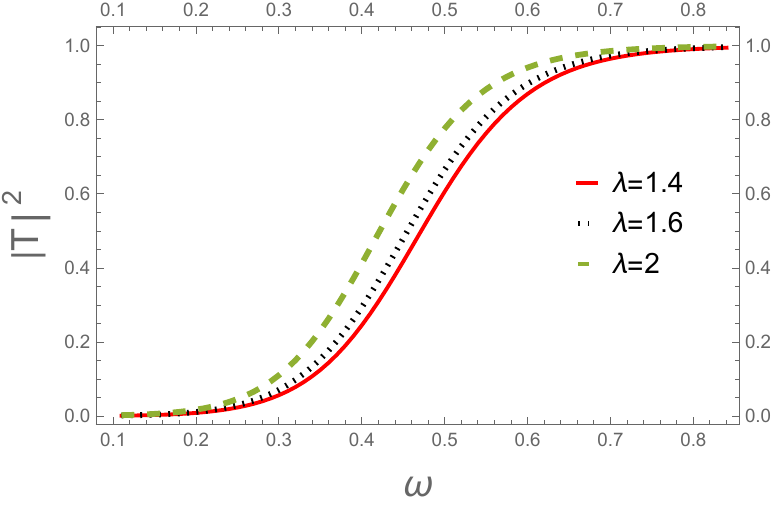}}
\subfigure[$a=-0.5$,$Q=0.4$]{\label{fig3g}%% label for first subfigure
\includegraphics[width=1.5in]{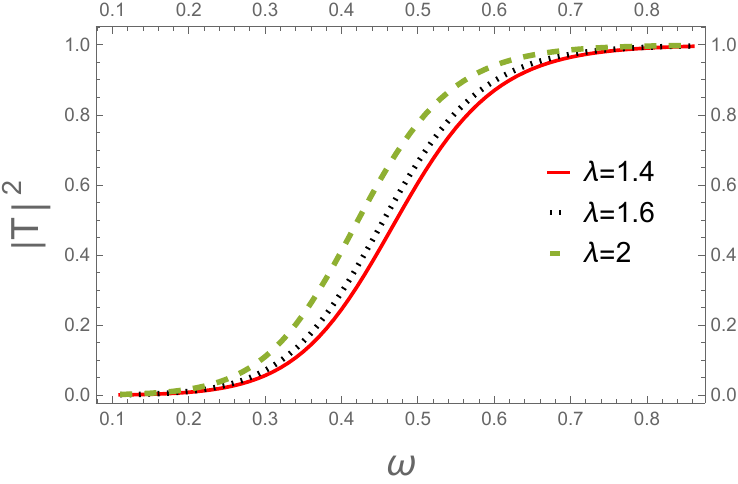}}
\caption{The scalar field greybody factor \textit{vs.} $\lambda$ with
$M=1$,$l=1$.}\label{figs1}
\end{figure}

\begin{figure}[H]
\centering
\subfigure[$a=0.5$,$Q=0.4$]{\label{fig1g}%% label for first subfigure
\includegraphics[width=1.5in]{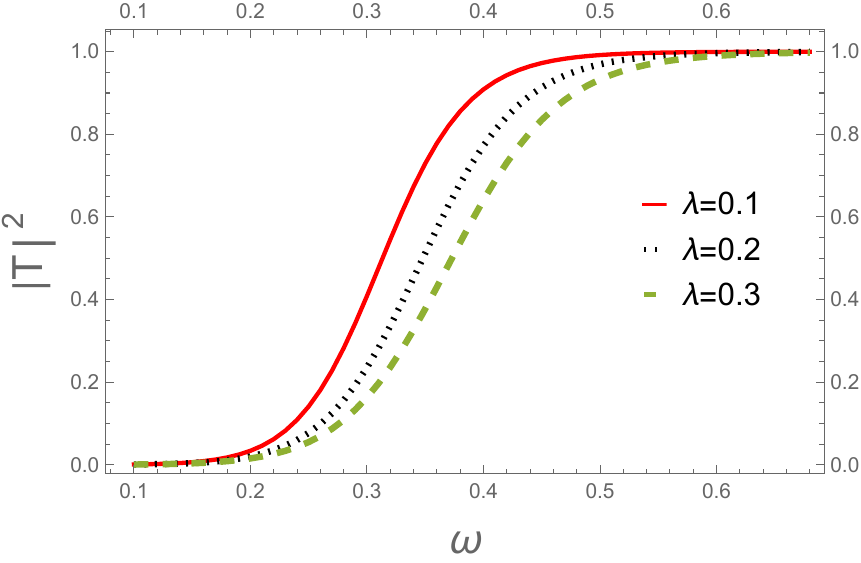}}
\subfigure[$a=-0.5$,$Q=0.4$]{\label{fig3g}%% label for first subfigure
\includegraphics[width=1.5in]{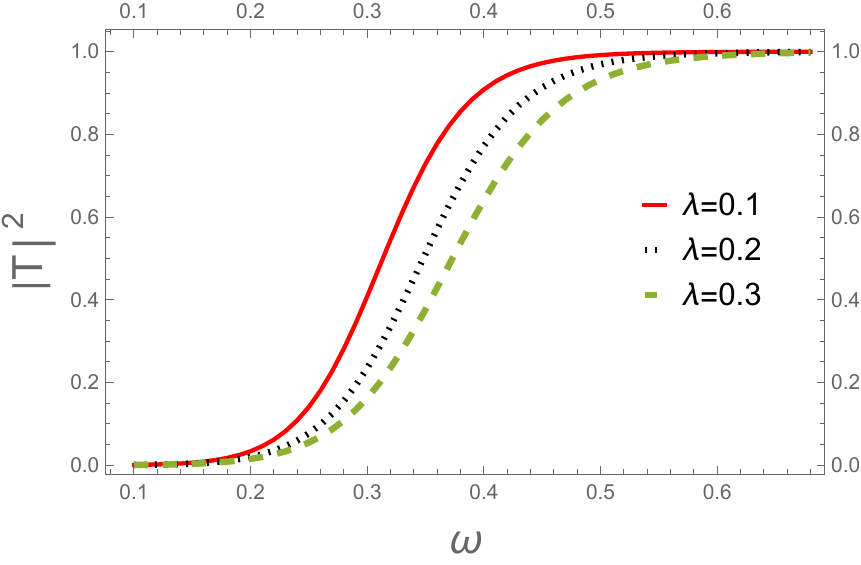}}
\subfigure[$a=0.5$,$Q=0.4$]{\label{fig1g}%% label for first subfigure
\includegraphics[width=1.5in]{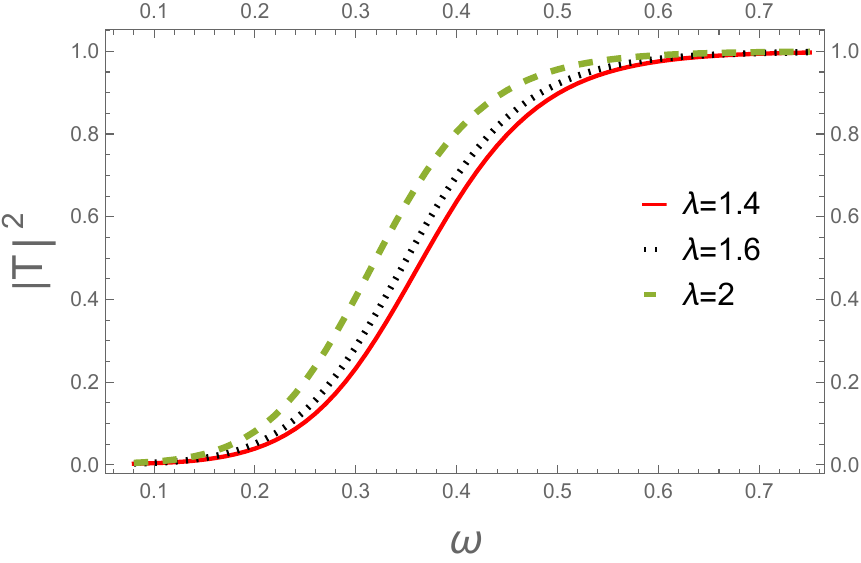}}
\subfigure[$a=-0.5$,$Q=0.4$]{\label{fig3g}%% label for first subfigure
\includegraphics[width=1.5in]{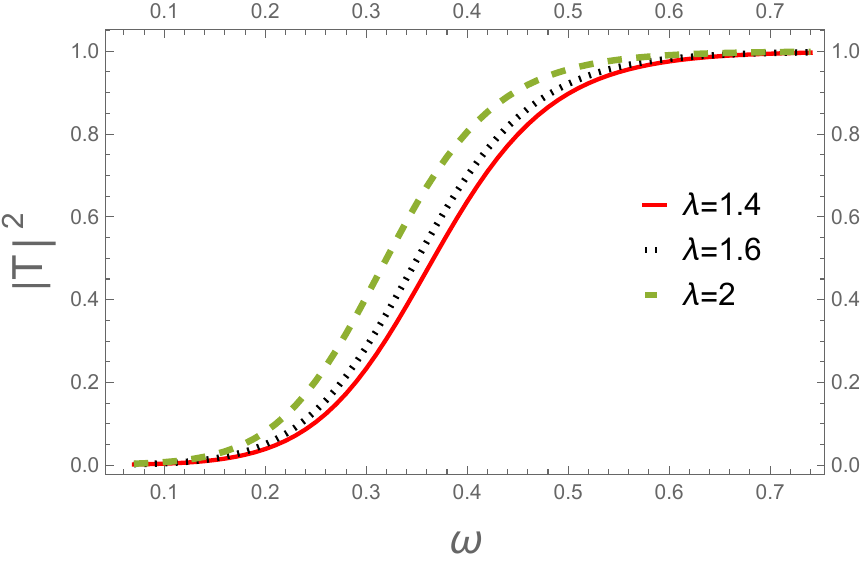}}
\caption{The electromagnetic field greybody factor \textit{vs.} $\lambda$ with  $M=1$,$l=1$.}\label{figs1}
\end{figure}

\section{Conclusion and discussion}
\label{sec5}

In this work, we have systematically investigated the quasinormal modes (QNMs) and greybody factors of massless scalar and electromagnetic perturbations around the Euler--Heisenberg black hole surrounded by perfect fluid dark matter (PFDM). Using the asymptotic iteration method (AIM) together with the sixth-order WKB approximation, we analyze the effects of the black hole charge $Q$, the PFDM parameter $\lambda$, and the nonlinear Euler--Heisenberg parameter $a$ on the dynamical properties of the spacetime. The excellent agreement between the AIM and WKB results confirms the reliability and accuracy of the obtained quasinormal spectra.

The effective potential analysis shows that both the black hole charge and the PFDM environment significantly modify the structure of the scattering barrier. Increasing the charge parameter $Q$ generally enhances the height of the effective potential barrier for both scalar and electromagnetic perturbations, leading to stronger confinement of perturbative waves near the photon-sphere region. By contrast, the PFDM parameter $\lambda$ produces a more intricate influence on the effective potential. In the weak-PFDM regime, increasing $\lambda$ raises and sharpens the effective potential barrier, whereas for sufficiently large values of $\lambda$ the barrier becomes lower and smoother, reducing the curvature near the potential peak. The nonlinear Euler--Heisenberg correction mainly affects the near-horizon geometry and becomes prominent in the strong nonlinear regime.

The quasinormal spectra exhibit several remarkable dynamical features. For moderate values of the nonlinear parameter $a$, the oscillation frequency $\omega_R$ increases monotonically with $Q$, while the damping magnitude $|\omega_I|$ displays a characteristic nonmonotonic dependence, reaching a clear minimum at the critical charge $Q_c$, the extreme parameter. This behavior originates from the competition between the height and curvature of the effective potential near its peak. The PFDM parameter $\lambda$ elevates $Q_c$ and sufficiently strong dark matter can even eliminate the nonmonotonic damping behavior entirely. These results indicate that environmental dark matter effects can qualitatively alter the ringdown properties of charged black holes.

A comparison between scalar and electromagnetic perturbations reveals that scalar modes are generally more sensitive to the background geometry and the PFDM environment. This distinction arises because the scalar effective potential contains an extra derivative term $rf(r)f'(r)$, whereas the electromagnetic potential depends only on the centrifugal contribution. Consequently, scalar perturbations exhibit stronger variations in both oscillation frequency and damping rate under changes of spacetime parameters. Nevertheless, both fields share consistent qualitative behavior, demonstrating the universality of the underlying physical mechanism.

In the strong nonlinear regime, the Euler--Heisenberg correction introduces qualitatively new effects into the quasinormal dynamics. The dependence of $\omega_R$ on $Q$ gradually changes from monotonic increase to a trend of first increasing and then decreasing near extremality. Meanwhile, the damping structure becomes more complex, indicating that strong nonlinear electrodynamic effects substantially reshape the curvature of the effective potential near its peak. A robust result is that the critical charge $Q_c$ increases monotonically with $a$, demonstrating that stronger nonlinear corrections push the extremal condition toward larger black hole charge.

The greybody-factor analysis further confirms the close connection between scattering properties and the effective potential structure. Increasing $Q$ suppresses low-frequency transmission, while $\lambda$ regulates wave confinement and tunneling. Both nonlinear electrodynamics and PFDM leave characteristic imprints on black hole absorption and scattering, which may be detectable in future observations.

Overall, the interplay between black hole charge, nonlinear electrodynamics, and perfect fluid dark matter yields rich and nontrivial modifications to quasinormal spectra and greybody factors. The competition between the height and curvature of the effective potential governs the oscillation and damping characteristics of perturbative modes. Our results provide valuable theoretical templates for future gravitational-wave observations, black hole spectroscopy, and tests of strong-field gravity and nonlinear electrodynamics in dark matter environments.

We remark that a related investigation on greybody factors was presented in Ref.~\cite{Belchior:2026}, while a recent study focusing on the eikonal approximation, shadow, and eikonal greybody factors in the same spacetime was reported in Ref.~\cite{Silva:2026cue}. Our work is independently conducted and provides a systematic study of more complete quasinormal mode spectra for both scalar and electromagnetic perturbations, together with greybody factors calculated beyond the eikonal limit, which extends and complements the results in the literature.

%Taking different multipole moment $l$, we found the QNFs obtained through the AIM method are in good agreement with those of obtained by the WKB method. Moreover, the PWKB method improves the precision of the fundamental QNFs. In addition, w

\appendix
\section{Asymptotic Iteration Method}
The asymptotic iteration method (AIM) provides a stable and accurate semi-analytical strategy for solving second-order linear ordinary differential equations that describe black hole perturbations~\cite{Cho:2009cj,Ciftci:2003}.
This appendix outlines the complete implementation of AIM for computing quasinormal frequencies, with all key equations preserved in their original form.

To begin, we introduce a compactified radial coordinate
\begin{equation}
u=1-\frac{r_{+}}{r},
\end{equation}
which maps the exterior region $r\in[r_{+},\infty)$ to $u\in[0,1)$.
Substituting this transformation into the radial perturbation equation yields the following second-order ordinary differential equation in terms of $u$:
\begin{equation}
\psi''(u)+\left(\frac{2}{u-1}+\frac{f'(u)}{f(u)}\right) \psi'(u)
+\left[\frac{r_{+}^{2} \omega^{2}+(u-1)^{3} f(u) f'(u)}{(u-1)^{4} f(u)^{2}}-\frac{l(l+1)}{(u-1)^{2} f(u)}\right] \psi(u)=0.
\tag{A2}
\end{equation}
This second-order differential equation governs the radial behavior of perturbations and serves as the starting point for the AIM procedure.

We first analyze the asymptotic behavior near the horizon $u\to0$.
The metric function can be expanded linearly as $f(u)\approx u f'(0)$.
Inserting this into (A1) gives the simplified equation
\begin{equation}
\psi''(u)+\frac{1}{u} \psi'(u)+\frac{r_{+}^{2} \omega^{2}}{u^{2} f'(0)^{2}} \psi(u)=0,
\tag{A3}
\end{equation}
whose general solution is
\begin{equation}
\psi(u\to0) \sim C_{1} u^{-\xi}+C_{2} u^{\xi},\quad \xi=\frac{i r_{+} \omega}{f'(0)}.
\tag{A4}
\end{equation}
The physical ingoing boundary condition requires $C_{2}=0$.

Next, we consider the asymptotic form at spatial infinity $u\to1$.
In this limit, the equation reduces to
\begin{equation}
\psi''(u)-\frac{2}{1-u} \psi'(u)+\frac{r_{+}^{2} \omega^{2}}{(1-u)^{4}} \psi(u)=0.
\tag{A5}
\end{equation}
By introducing $\zeta=\frac{i r_{+}\omega}{1-u}$, the solution becomes
\begin{equation}
\psi(u\to1)\sim D_{1} e^{-\zeta}+D_{2} e^{\zeta}.
\tag{A6}
\end{equation}
The purely outgoing condition selects $D_{1}=0$.

We now construct the global ansatz that satisfies both boundary conditions:
\begin{equation}
\psi(u)=u^{-\xi}e^{\zeta}\chi(u),
\tag{A7}
\end{equation}
where $\chi(u)$ is regular on $(0,1)$.
Substituting (A6) into (A1) casts the equation into the canonical AIM form~\cite{Ciftci:2003,Cho:2009cj}:
\begin{equation}
\chi''=\lambda_{0}(u)\chi'+s_{0}(u)\chi.
\tag{A8}
\end{equation}

For scalar perturbations, the initial iteration coefficients are given by
\begin{equation}
\lambda_{0}(u)=\frac{2i r_{+}\omega}{u f'(0)}-\frac{f'(u)}{f(u)}-\frac{2\left(i r_{+}\omega+u-1\right)}{(u-1)^{2}},
\tag{A9}
\end{equation}
and
\begin{align}
s_{0}(u) &= \frac{1}{(u-1)^{4}u^{2}f(u)^{2}f'(0)^{2}}\Bigl[ i r_{+}(u-1)^{3}(u+1)\omega f'(0)f(u)^{2} \nonumber\\
&\quad -u^{2}f'(0)^{2}\bigl(r_{+}^{2}\omega^{2}+(u-1)^{2}f(u)f'(u)(i r_{+}\omega+u-1)\bigr) \nonumber\\
&\quad +f(u)\bigl(l(l+1)u^{2}(u-1)^{2}f'(0)^{2}+i r_{+}u(u-1)^{4}\omega f'(0)f'(u) \nonumber\\
&\quad +r_{+}^{2}\omega^{2}f(u)\left((u-1)^{2}-u f'(0)\right)^{2}\bigr)\Bigr].
\tag{A10}
\end{align}

For electromagnetic perturbations, the coefficient $\lambda_{0}(u)$ remains identical:
\begin{equation}
\lambda_{0}(u)=\frac{2i r_{+}\omega}{u f'(0)}-\frac{f'(u)}{f(u)}-\frac{2\left(i r_{+}\omega+u-1\right)}{(u-1)^{2}},
\tag{A11}
\end{equation}
while $s_{0}(u)$ takes the form
\begin{align}
s_{0}(u) &= \frac{1}{(u-1)^{4}u^{2}f(u)^{2}f'(0)^{2}}\Bigl[ -r_{+}^{2}u^{2}\omega^{2}f'(0)^{2} \nonumber\\
&\quad +r_{+}\omega f(u)^{2}\bigl(r_{+}\omega\left((u-1)^{2}-u f'(0)\right)^{2}+i(u-1)^{3}(1+u)f'(0)\bigr) \nonumber\\
&\quad +l(1+l)(u-1)^{2}u^{2}f(u)f'(0)^{2}+i r_{+}(u-1)^{2}u\omega f'(0)\bigl((u-1)^{2}-u f'(0)\bigr)f(u)f'(u)\Bigr].
\tag{A12}
\end{align}

The iteration then proceeds recursively to convergence, yielding robust and high-precision quasinormal frequencies.

\section{6th-order WKB method}
The sixth-order Wentzel--Kramers--Brillouin (WKB) approximation serves as a reliable semi-analytic tool for computing quasinormal frequencies and greybody factors in black hole perturbation theory~\cite{Konoplya:2003ii}.
This approach approximates the perturbed field equation around the peak of the effective potential barrier, where the field exhibits the dominant oscillatory behavior.

The central condition determining the quasinormal frequencies takes the form
\begin{equation}
\frac{i\left(\omega^2-V_0\right)}{\sqrt{-2V_0''}}
-\sum_{k=2}^6\Lambda_k
= n+\frac{1}{2},
\end{equation}
where \(n=0,1,2,\dots\) labels the overtone number.
Here \(V_0=V(r_0)\) denotes the peak value of the effective potential at its maximum \(r=r_0\),
and \(V_0''\) stands for the second derivative at the peak, characterizing the curvature of the potential barrier.
The terms \(\Lambda_k\) represent high-order corrections derived from the \(k\)-th derivatives of \(V(r)\) at \(r_0\),
which systematically improve the accuracy up to the sixth order.

Unlike low-order expansions, the sixth-order scheme reliably captures the potential's local structure near the peak,
yielding accurate frequencies even for moderately strong-field regimes~\cite{Konoplya:2003ii}.
This method is computationally efficient and consistent with the asymptotic iteration method~\cite{Cho:2009cj},
especially for modes satisfying \(l\ge n\), for which the potential barrier is sufficiently pronounced and the WKB expansion converges rapidly.
The same framework is employed to compute greybody factors by evaluating the transmission probability across the potential barrier.


\begin{thebibliography}{99}
%\cite{Abbott:2016blz}
\bibitem{Abbott:2016blz}
B.~P.~Abbott \textit{et al.} [LIGO Scientific and Virgo Collaborations],
%``Observation of Gravitational Waves from a Binary Black Hole Merger,''
Phys.\ Rev.\ Lett.\ \textbf{116} (2016) no.6, 061102
[arXiv:1602.03837 [gr-qc]].

%\cite{Abbott:2017oio}
\bibitem{Abbott:2017oio}
B.~P.~Abbott \textit{et al.} [LIGO Scientific and Virgo Collaborations],
%``GW170817: Observation of Gravitational Waves from a Binary Neutron Star Inspiral,''
Phys.\ Rev.\ Lett.\ \textbf{119} (2017) no.16, 161101
[arXiv:1710.05832 [gr-qc]].

%\cite{Kokkotas:1999bd}
\bibitem{Kokkotas:1999bd}
K.~D.~Kokkotas and B.~G.~Schmidt,
%``Quasinormal modes of stars and black holes,''
Living Rev.\ Rel.\ \textbf{2} (1999), 2
[arXiv:gr-qc/9909058 [gr-qc]].

%\cite{Berti:2009kk}
\bibitem{Berti:2009kk}
E.~Berti, V.~Cardoso and A.~O.~Starinets,
%``Quasinormal modes of black holes and black branes,''
Class.\ Quant.\ Grav.\ \textbf{26} (2009), 163001
[arXiv:0905.2975 [gr-qc]].

%\cite{Konoplya:2011qq}
\bibitem{Konoplya:2011qq}
R.~A.~Konoplya and A.~Zhidenko,
%``Quasinormal modes of black holes: From astrophysics to string theory,''
Rev.\ Mod.\ Phys.\ \textbf{83} (2011), 793-836
[arXiv:1102.4014 [gr-qc]].

%\cite{Cardoso:2019rvt}
\bibitem{Cardoso:2019rvt}
V.~Cardoso and P.~Pani,
%``Testing the nature of dark compact objects: a status report,''
Living Rev.\ Rel.\ \textbf{22} (2019) no.1, 4
[arXiv:1904.05363 [gr-qc]].

%\cite{Berti:2018vdi}
\bibitem{Berti:2018vdi}
E.~Berti, K.~Yagi and N.~Yunes,
%``Extreme Gravity Tests with Gravitational Waves from Compact Binary Coalescences: (I) Inspiral-Merger,''
Gen.\ Rel.\ Grav.\ \textbf{50} (2018) no.4, 46
[arXiv:1801.03208 [gr-qc]].

%\cite{Rubin:1980zd}
\bibitem{Rubin:1980zd}
V.~C.~Rubin, N.~Thonnard and W.~K.~Ford,
%``Rotational properties of 21 SC galaxies with a large range of luminosities and radii,''
Astrophys.\ J.\ \textbf{238} (1980), 471-487.

%\cite{Clowe:2006eq}
\bibitem{Clowe:2006eq}
D.~Clowe, M.~Brada\v{c}, A.~H.~Gonzalez, M.~Markevitch, S.~W.~Randall, C.~Jones and D.~Zaritsky,
%``A direct empirical proof of the existence of dark matter,''
Astrophys.\ J.\ Lett.\ \textbf{648} (2006), L109-L113
[arXiv:astro-ph/0608407 [astro-ph]].

%\cite{Aghanim:2018eyx}
\bibitem{Aghanim:2018eyx}
N.~Aghanim \textit{et al.} [Planck Collaboration],
%``Planck 2018 results. VI. Cosmological parameters,''
Astron.\ Astrophys.\ \textbf{641} (2020), A6
[arXiv:1807.06209 [astro-ph.CO]].

%\cite{Li:2013fka}
\bibitem{Li:2013fka}
M.~Li, K.~Yang and Y.~Zhong,
%``Exact black hole solutions and noncommutative effects in dark matter background,''
JCAP \textbf{09} (2013), 043
[arXiv:1307.4658 [gr-qc]].

%\cite{Xu:2018}
\bibitem{Xu:2018}
Z.~Xu, X.~Hou, X.~Gong and J.~Wang,
%``Black hole space-time in dark matter halo,''
JCAP \textbf{09} (2018), 038
[arXiv:1803.00767 [gr-qc]].

%\cite{Euler:1935zz}
\bibitem{Euler:1935zz}
H.~Euler and B.~Kockel,
%``The scattering of light by light in Dirac's theory,''
Naturwiss.\ \textbf{23} (1935), 246-247.

%\cite{Heisenberg:1936nmg}
\bibitem{Heisenberg:1936nmg}
W.~Heisenberg and H.~Euler,
%``Consequences of Dirac's theory of positrons,''
Z.\ Phys.\ \textbf{98} (1936) no.11-12, 714-732
[arXiv:physics/0605038 [physics]].

%\cite{Kruglov:2017fut}
\bibitem{Kruglov:2017fut}
S.~I.~Kruglov,
%``Black hole as a magnetic monopole within exponential nonlinear electrodynamics,''
Annals Phys.\ \textbf{378} (2017), 59-70
[arXiv:1703.02174 [physics.gen-ph]].

%\cite{Rodrigues:2023}
\bibitem{Rodrigues:2023}
M.~E.~Rodrigues and M.~V.~de~Sousa,
%``Optical properties of nonlinear electrodynamics black holes,''
Phys.\ Dark Univ.\ \textbf{41} (2023), 101255.

%\cite{Ma:2024}
\bibitem{Ma:2024}
Y.~Ma, H.~Wu and Y.~Yang,
%``Euler--Heisenberg black holes surrounded by perfect fluid dark matter,''
Phys.\ Dark Univ.\ \textbf{45} (2024), 101234.

%\cite{Su:2024}
\bibitem{Su:2024}
Y.~Su and H.~Qiao,
%``Black hole solutions in Euler--Heisenberg electrodynamics with perfect fluid dark matter,''
Eur.\ Phys.\ J.\ C \textbf{84} (2024), 456.



%\cite{Akiyama:2019cqa}
\bibitem{Akiyama:2019cqa}
K.~Akiyama \textit{et al.} [Event Horizon Telescope Collaboration],
%``First M87 Event Horizon Telescope Results. I. The Shadow of the Supermassive Black Hole,''
Astrophys.\ J.\ Lett.\ \textbf{875} (2019), L1
[arXiv:1906.11238 [astro-ph.GA]].

%\cite{Akiyama:2022}
\bibitem{Akiyama:2022}
K.~Akiyama \textit{et al.} [Event Horizon Telescope Collaboration],
%``First Sagittarius A* Event Horizon Telescope Results. I. The Shadow of the Supermassive Black Hole,''
Astrophys.\ J.\ Lett.\ \textbf{930} (2022), L12
[arXiv:2205.11579 [astro-ph.GA]].

%\cite{Cho:2009cj}
\bibitem{Cho:2009cj}
H.~T.~Cho, A.~S.~Cornell, J.~Doukas and W.~Naylor,
%``Black hole quasinormal modes using the asymptotic iteration method,''
Class.\ Quant.\ Grav.\ \textbf{27} (2010), 155004
[arXiv:0912.2740 [gr-qc]].

%\cite{Konoplya:2003ii}
\bibitem{Konoplya:2003ii}
R.~A.~Konoplya,
%``Quasinormal behavior of the d-dimensional Schwarzschild black hole and higher order WKB approach,''
Phys.\ Rev.\ D \textbf{68} (2003), 024018
[arXiv:gr-qc/0303052 [gr-qc]].

%\cite{Li:2013}
\bibitem{Li:2013}
M.~Li, K.~Yang and Y.~Zhong,
JCAP \textbf{09} (2013), 043
[arXiv:1307.4658 [gr-qc]].

%\cite{Xu:2018}
\bibitem{Xu:2018}
Z.~Xu, X.~Hou, X.~Gong and J.~Wang,
JCAP \textbf{09} (2018), 038
[arXiv:1803.00767 [gr-qc]].

%\cite{Konoplya:2025ect}
\bibitem{Konoplya:2025ect}
R.~A.~Konoplya and A.~Zhidenko,
Phys.~Rev.~D \textbf{113} (2026) 043011.

%\cite{Konoplya:2022hbl}
\bibitem{Konoplya:2022hbl}
R.~A.~Konoplya and A.~Zhidenko,
Astrophys.~J. \textbf{933} (2022) 166.

%\cite{Ma:2024}
\bibitem{Ma:2024}
Y.~Ma, H.~Wu and Y.~Yang,
Phys.~Dark Univ. \textbf{45} (2024), 101234.

%\cite{Su:2024}
\bibitem{Su:2024}
Y.~Su and H.~Qiao,
Eur.~Phys.~J.~C \textbf{84} (2024), 456.

%\cite{Euler:1935}
\bibitem{Euler:1935}
H.~Euler and B.~Kockel,
Naturwiss. \textbf{23} (1935), 246--247.

%\cite{Heisenberg:1936}
\bibitem{Heisenberg:1936}
W.~Heisenberg and H.~Euler,
Z.~Phys. \textbf{98} (1936), 714--732.

%\cite{Kruglov:2017}
\bibitem{Kruglov:2017}
S.~I.~Kruglov,
Annals Phys. \textbf{378} (2017), 59--70
[arXiv:1703.02174 [physics.gen-ph]].

%\cite{Schwarzschild:1916}
\bibitem{Schwarzschild:1916}
K.~Schwarzschild,
Sitzungsber.~Preuss.~Akad.~Wiss.~Berlin (1916), 189--196.

%\cite{Reissner:1916}
\bibitem{Reissner:1916}
H.~Reissner,
Ann.~Phys. \textbf{355} (1916), 106--120.

%\cite{Nordstrom:1918}
\bibitem{Nordstrom:1918}
G.~Nordström,
Proc.~Kon.~Ned.~Akad.~Wet. \textbf{20} (1918), 1238--1245.

%\cite{Regge:1957td}
\bibitem{Regge:1957td}
T.~Regge and J.~A.~Wheeler,
Stability of a Schwarzschild singularity,
Phys.\ Rev.\ \textbf{108} (1957), 1063--1069.


%\cite{Vishveshwara:1970zz}
\bibitem{Vishveshwara:1970zz}
C.~V.~Vishveshwara,
Scattering of gravitational radiation by a Schwarzschild black-hole,
Nature \textbf{227} (1970), 936--938.



%\cite{Iyer:1986np}
\bibitem{Iyer:1986np}
S.~Iyer and C.~M.~Will,
%``Black hole normal modes: A WKB approach. 1. Foundations and application of a higher order WKB analysis of potential barrier scattering,''
Phys.\ Rev.\ D \textbf{35} (1987), 3621.

%\cite{Page:1976df}
\bibitem{Page:1976df}
D.~N.~Page,
%``Particle emission rates from a black hole: Massless particles from an uncharged, nonrotating hole,''
Phys.\ Rev.\ D \textbf{13} (1976), 198-206.

%\cite{Harmark:2007jy}
\bibitem{Harmark:2007jy}
T.~Harmark, J.~Natario and R.~Schiappa,
%``Greybody Factors for d-Dimensional Black Holes,''
Adv.\ Theor.\ Math.\ Phys.\ \textbf{14} (2010), 727-794
[arXiv:0708.0017 [hep-th]].


%\cite{Konoplya:2021ube}
\bibitem{Konoplya:2021ube}
R.~A.~Konoplya,
Phys.~Lett.~B \textbf{823} (2021) 136734.


%\cite{Ciftci:2003}
\bibitem{Ciftci:2003}
H.~Ciftci, R.~L.~Hall and N.~Saad,
%``Asymptotic iteration method for eigenvalue problems,''
J.\ Phys.\ A \textbf{36} (2003), 11807-11816
[arXiv:math-ph/0309064 [math-ph]].





%\cite{Belchior:2026}
\bibitem{Belchior:2026}
H.~S.~Belchior \textit{et al.},
%``Greybody factors and absorption cross sections of Euler--Heisenberg black holes surrounded by perfect fluid dark matter,''
arXiv:2605.04994 [gr-qc].

%\cite{Silva:2026cue}
\bibitem{Silva:2026cue}
E.~O.~Silva and F.~Ahmed,
``Shadow, quasinormal modes, sparsity, and energy emission rate of Euler–Heisenberg black hole surrounded by perfect fluid dark matter,''
[arXiv:2604.16628 [gr-qc]].



\end{thebibliography}
\end{document}